\documentclass[letterpaper]{ar-1col}

\usepackage[comma]{natbib}
\usepackage{url}

\jname{Xxxx. Xxx. Xxx. Xxx.}
\jvol{AA}
\jyear{YYYY}
\doi{10.1146/((please add article doi))}

\def\adhoc{{\it ad hoc}}
\def\cf{{\it cf.}}
\def\eg{{\it e.g.}}
\def\etal{{\it et al.}}
\def\etc{{\it etc.}}
\def\ie{{\it i.e.}}

\def\Gaia{{\it Gaia\/}}

\def\spose#1{\hbox to 0pt{#1\hss}} % from Scott Tremaine
\def\ltsim{\mathrel{\spose{\lower.5ex\hbox{$\mathchar"218$}}
     \raise.4ex\hbox{$\mathchar"13C$}}}
\def\gtsim{\mathrel{\spose{\lower.5ex \hbox{$\mathchar"218$}}
     \raise.4ex\hbox{$\mathchar"13E$}}}
\def\gtlt{\mathrel{\spose{\lower.5ex\hbox{$\mathchar"13E$}}
     \raise.5ex\hbox{$\mathchar"13C$}}}
\def\pmb#1{\setbox0=\hbox{$#1$}%
  \kern-0.25em\copy0\kern-\wd0
  \kern.05em\copy0\kern-\wd0
  \kern-0.025em\raise.0433em\box0}
\def\spmb#1{\setbox1=\hbox{${\scriptstyle #1}$}%
  \kern-0.25em\copy1\kern-\wd1
  \kern.05em\copy1\kern-\wd1
  \kern-0.025em\raise.0433em\box1}

\long\def\Ignore#1{\relax}
\usepackage{color}
\definecolor{red}{rgb}{0.7,0.1,0.1}
\definecolor{blue}{rgb}{0.0,0.0,1.0}
\definecolor{green}{rgb}{0.1,0.6,0.1}

\begin{document}

% Page header
\markboth{Sellwood and Masters}{Spirals in galaxies}

% Title
\title{Spirals in galaxies}

%Authors, affiliations address.
\author{J. A. Sellwood$^1$ and Karen L. Masters$^2$
\affil{$^1$Steward Observatory, University of Arizona, Tucson, Arizona, USA; email: sellwood@as.arizona.edu}
\affil{$^2$Departments of Physics and Astronomy, Haverford College, Haverford, Pennsylvania, USA; email: klmasters@haverford.edu}}

%Abstract
\begin{abstract}
Spirals in galaxies have long been thought to be caused by
gravitational instability in the stellar component of the disk, but
discerning the precise mechanism had proved elusive.  Tidal
interactions, and perhaps bars, may provoke some spiral responses, but
spirals in many galaxies must be self-excited.  We survey the relevant
observational data and aspects of disk dynamical theory.  The origin
of the recurring spiral patterns in simulations of isolated disk
galaxies has recently become clear and it is likely that the mechanism
is the same in real galaxies, although evidence to confirm this
supposition is hard to obtain.  As transient spiral activity increases
random motion, the patterns must fade over time unless the disk also
contains a dissipative gas component.  Continuing spiral activity
alters the structure of the disks in other ways: reducing metallicity
gradients and flattening rotation curves are two of the most
significant.  The overwhelming majority of spirals in galaxies have
two- or three-fold rotational symmetry, indicating that the cool, thin
disk component is massive.  Spirals in simulations of halo-dominated
disks instead manifest many arms, and consequently do not capture the
expected full spiral-driven evolution.  We conclude by identifying
areas where further work is needed.

To appear in vol 60 of ARAA
\end{abstract}

%Keywords, etc.
\maketitle

%Table of Contents
\tableofcontents

\section{INTRODUCTION}
The question ``why do large disk galaxies manifest spirals?''\ has a
simple, though rather unsatisfying, answer.  Evolution in any
rotationally supported disk, be it an accretion disk, a disk in a
proto-planetary system, a planetary ring system, or a disk galaxy, is
driven by outward angular momentum transport.  Some viscous-like
property is required in shearing disks to extract angular momentum
from the inner parts and deposit it farther out.  Because a
collisionless disk of stars is inviscid, outward transport of angular
momentum is accomplished by the gravitational stress from massive,
trailing spiral features.  However, this explanation does not begin to
address the much harder question that is the topic of this review:
{\em how exactly} are spiral patterns created in disk galaxies?

\Ignore{Approximately 175 years ago,} Lord Rosse was the first to
observe spiral arms in a galaxy; he published \citep{Rosse1846} a
sketch of the pattern in Messier 51, \ie\ the Whirlpool Galaxy.
However, it took many decades before spiral nebulae were recognized as
external galaxies having sizes comparable to that of the Milky Way,
and still longer to discover that the larger galaxies had flat
rotation curves, and even now the extent to which the central
attraction is dominated by the stars and gas in the disk is still
debated \citep[\eg][]{Post19, Macc20}.

The prominence of young stars and HII regions at first suggested that
spirals were a gas-dynamical phenomenon, perhaps controlled by
magnetic fields.  However, Bertil Lindblad (pronounced ``Lindblard'')
in the 1940s and 50s, suggested that self-gravity of the stars could
be important, and the idea was taken up in earnest by the applied
mathematicians of MIT and Harvard University in the early 1960s.  One
of the first conferences JAS attended as a graduate student was IAU
Symp.\ 77, where he witnessed an intense discussion about the cause of
spiral patterns and resolution of the issues debated seemed far off.
\citet{Pasha02, Pasha04} gives a well-researched historical account of
the developing clash of ideas up to about this date.

By the early 1970s, computers had become powerful enough to follow the
collective dynamics of many thousands of self-gravitating particles.
The first simulations \citep*{MPQ70, Hohl71, HB74} revealed that
models bearing some resemblance to disk galaxies could manifest spiral
features for a short time, seeming to confirm that they had captured
the essential physics, but it has taken almost another half-century to
understand the mechanism that causes spiral instabilities in
simulations of isolated stellar disks \citep{SC19}.  Although we
review this progress, it should be noted at the outset that we still
lack truly compelling evidence that any spirals in real galaxies are
caused by the mechanism that has been identified from the simulations.

Furthermore, it has been established that spirals in galaxies need not
be exclusively self-excited instabilities of an equilibrium disk, but
they can be the driven responses to passing companions, substructure
in the halo, and perhaps also the stellar bar that resides in the
inner parts of a large fraction of disk galaxies.  The spiral
responses of the disk to these driving mechanisms may seem to beg the
question: do we need a self-excitation mechanism at all?  We review
the evidence that one is needed in \S\ref{sec.driven}.

Simulations have also revealed that spirals are important drivers of
secular evolution in galaxies, and that the present-day structure of
disk galaxies is not simply the result of initial conditions at the
time of formation \citep[\eg][]{KK04}.  Spiral-driven evolution can
alter the distribution of angular momentum within the disk, contribute
to the increased random motion of older disk stars, smooth rotation
curves, assist galactic dynamos, and cause a widespread diffusion of
the orbit radii of stars, with important consequences for the spatial
and age distributions of metals among the stars.

\citet{To77} provided an insightful, though now rather dated, review
of spiral structure.\Ignore{, that is still worth reading despite
  being rather dated, was provided by} Among other reviews,
\citet{Athan84} had a similar interpretation, but \citet{DB14} and
\citet{Shu16} each embraced radically differing perspectives.
\citet{BT08} provide a clear introduction to some of the basic
concepts and mathematical derivations of important formulae, which we
will rely upon in this review.

\section{OBSERVED PROPERTIES OF SPIRALS}
\label{sec.obs}
Setting aside dwarfs, the majority of galaxies in the local Universe
are spiral galaxies.  This fact noticed in early galaxy surveys
continues to hold in the {\it Galaxy Zoo} crowd-sourced morphologies
of close to a million galaxies \citep[\eg][]{Lintott2011,Willett13}.

We here give a brief review of the general nature of spiral arms in
galaxies, but our goal is to focus on properties of the patterns, such
as arm multiplicity, pitch angle, arm amplitude, \etc, that could
provide some constraints on models.  Several theories for the origin
of spiral patterns, reviewed in \S\ref{sec.theories}, invoke strong
disk responses and/or swing-amplification (see \S\ref{sec.responses})
and therefore make virtually identical predictions for most
observables, even though each such theory proposes a different origin
for what is being amplified.  However, ``quasi-stationary density waves''
\citep[\S\ref{sec.waser},][p9]{BL96} are mildly amplified in general,
and therefore the observed properties of the spiral patterns are not
expected to conform with the predictions of swing-amplification in
this case.

\subsection{Modern Demographics of Spiral Arms in Galaxies}
\label{sec.obs-gen}
The Main Galaxy Survey \citep{Strauss2002} of the Sloan Digital Sky
Survey, completed over a decade ago, provided well-resolved images in
five color bands of almost $250\,000$ galaxies.  Yet progress on
quantifying spiral structure has been slow because measurements of the
spiral patterns present a particular challenge.  Automated, or
semi-automated, photometric decompositions generally ignore, or
azimuthally average, the spiral features, and most detailed
observational studies of spiral arms have been made in relatively
modest samples, necessitated by visual inspection, or other intensive
image analysis.

Spiral patterns are complex structures, whose morpologies can differ
in a variety of ways, including amplitude (\ie\ arm-inter-arm
contrast), width, pitch angle, number of arms, and
patchiness.\begin{marginnote}
\entry{Pitch Angle}{The angle, $\alpha$, in the disk plane that a
  spiral arm makes with a circle at the same radius.}
\end{marginnote}Hubble originally ordered spiral galaxies in a 
sequence as Sa, Sb or Sc\footnote{Intermediate types Sab, Sbc and S0,
  Sd and Sm galaxies were added later} based on a combination of the
prominence of the central bulge and the ``the spiral pitch angle and
degree of resoution'' of the spiral arms \citet[][p325]{Hubble26}.
Disk galaxies lacking any spiral patterns, gas, or dust were
classified as S0, or lenticular.  With no implication of an
evolutionary sequence, Hubble labeled disk galaxies to range from
S0/Sa ``early-type" to Sc/Sd ``late-type" spirals. The Hubble
classification correlates with galaxy color, with lenticulars and
ellipticals being redder, while spirals are increasingly blue towards
the later types.  Galaxies are bi-modal both in morphology and color,
but care should be taken with equating red with early-type and blue
with spiral as interesting sub-populations of red spirals and blue
early types do exist (\eg\ \citealt{Smeth22}).

The Hubble sequence does not capture all the ways in which spiral arms
vary in appearance.  \citet{ElEl82} coined the apt word ``flocculent"
to describe NGC~2841-type galaxies that had previously been identified
as a separate ``division'' by \citet{Sand61} and by \citet{KN79}.
\citet{ElEl82} went on to describe a set of arm-classes from``grand
design" to ``flocculent".  Grand design patterns generally, though not
exclusively, have just two arms and are slightly more common among the
earlier Hubble types,\begin{marginnote} \entry{Flocculent}{Literally
    having a fleece like appearance, meaning spiral arms that are
    short, closely spaced and fragmented} \entry{Grand design}{Spiral
    arms that are highly symmetric and continuous}
\end{marginnote}whereas the occurrence of flocculents increases toward
the later spirals.  But the trend is not strong; for example,
\citet{ElEl82} find a complete range from flocculent to grand design
among Sb-Sc galaxies.

The wavelength of observation strongly affects the appearance of
spiral arms.  As older stars contribute relatively more light to
redder bands, red and IR images better reflect the underlying stellar
population, while bright young stars stand out in images in bluer
bands.  As a consequence, spiral arms appear smoother in near-IR (NIR)
images \citep[\eg][]{Ja03}.  \citet{ElEl11} found from the S4G sample
\citep{Sheth10} that most optically flocculent galaxies are still at
least partially flocculent in the mid-IR (MIR), although some galaxies
which appear flocculent in blue light, are revealed to have underlying
grand design spirals in NIR images \citep{Th96}. Such examples are
rare, however, and \citet{Buta10} used the S4G survey to confirm the
earlier conclusion by \citet{Eskridge02} that classifications of most
spiral galaxies in the MIR are similar to those from B-band images,
being roughly one Hubble type earlier mainly because bulges become
more prominent.

One successful method of analyzing galaxy morphology in larger
samples, has been the use of crowd-sourcing (citizen science) to
obtain quantitative visual classifications \citep[\ie\ {\it Galaxy
    Zoo,}][]{Lintott2011}. Using {\it Galaxy Zoo} classifications, in
a series of papers, \citet{Hart2016,Hart2017a,Hart2017b,Hart2018}
provide a detailed look at the demographics and properties of over
6,000 galaxies having visible spiral arms in the redshift range
$0.03<z<0.05$ and with $r$-band magnitude $M_r<-21$ (or stellar masses
$\gtsim 10^{10}M_\odot$).

\subsection{Rotational Symmetry}
\label{sec.prefm}
Few spiral galaxies manifest highly regular and completely symmetric
spiral patterns, but a rough symmetry can usually be picked out by
eye.  \citet{Hart2016} found that the majority, 62\% in their
luminosity limited sample, of spiral galaxies have two spiral arms,
20\% of galaxies have three arms, 6.5\% have four arms, and a similar
percentage has five or more.  These proportions were found to depend
somewhat on galaxy properties, such that two-armed spirals are more
common in higher density environments and in redder disk galaxies.
These percentages from crowd-sourced visual inspection agree well
with findings from Fourier analysis of images of smaller samples
\citep[\eg][]{Davis12, YH18}.

The order of rotational symmetry of spiral patterns is predicted by
swing amplification theory (\S\ref{sec.swamp}) to correlate, albeit
with significant scatter, inversely with the disk contribution to the
central attraction \citep{SC84, ABP}.  A two or three arm pattern is
indicative of a heavy, almost maximum disk,\footnote{The contribution
  to the central attraction from an absolute maximum disk is as large
  as it can be without the rotation curve requiring a hollow halo.}
while significantly sub-maximum disks should manifest higher
rotational symmetry.  Furthermore, spiral patterns in galaxies
generally have higher multiplicity in the outer parts where the halo
contribution becomes more dominant.  If the large number of spiral
fragments in flocculent galaxies are gravitationally-driven, then that
would indicate a very low mass, cool, sub-component of the disk, which
may co-exist with an old unresponsive hot disk (see
\S\ref{sec.floccs}).

The review by \citet{JoCo09} reports that fully a third of spiral
galaxies exhibit significant asymmetry or lopsidedness.  Lopsided
instabilities are predicted in galaxy disks that lack any significant
halo \citep{Zang76, ER98b}, but this seems an unlikely explanation.
Therefore asymmetries are typically attributed \citep[\eg][]{ZR97} to
some kind of external forcing, such as tidal interactions or gas
inflow.

\subsection{Pitch Angle Measurements}
\label{sec.obs=PA}
Logarithmic spirals, which have a constant pitch angle with radius,
frequently fit spiral arms reasonably well \citep{Kennicutt81},
although individual arms can rarely be traced over a significant
radial range.  Measuring the pitch angle of an arm, let alone an
average over all arms in a galaxy, is complicated by the patchy nature
of spirals, kinks or branches that occur in some arms, differences
between pitch angles in multiple arms in a single galaxy, and the
uncertainty created by determining the inclination and center of the
galaxy.  All these factors mean that visual estimates of spiral
winding are easy, but quantitative measurement of pitch angles is
deceptively tricky \citep[see \eg][for recent comparisons of
  measurement techniques]{DG19,Hewitt20,YH20}.  These complications,
together with sample selection, have prevented the emergence of a
consensus on correlations between spiral arm pitch angle and global or
local galaxy properties.

The classic Hubble sequence implies a correlation between bulge size
and pitch angle, and such a correlation has been found, albeit with
large scatter \citep{Kennicutt81, Davis15, YH20}, while others have
reported an absence of a strong correlation \citep{Hart2017b,
  Masters19, DG19, Li21}.  Two recent papers have attempted to
correlate pitch angles with other galaxy properties in large samples
($N>2000$; \citealt{Hart2017b, YH20}), with both finding a greater
frequency of tightly wound arms in redder disks that have greater
stellar mass as well as galaxies with higher concentration and greater
stellar velocity dispersion.

Swing amplification predicts a loose correlation between pitch angle
and rotation curve slope: galaxies having rising rotation curves
should have more open arms while more tightly wrapped arms are
expected where rotation curves decline \citep[\S\ref{sec.swamp}
  and][]{Gran13}. A similar conclusion about the mass distribution was
reached by \citet{RRS75}.  \citet{Kennicutt81} and \citet{Seigar2006}
both noted a correlation between pitch angle and maximum circular
speed in the disk, but \citet{Seigar2006} linked the trend to the
impact of shear on pitch angle.  However, \citet{YH19} find a much
weaker correlation with shear, and they argue that the central slope
of the rotation curve is more important.

Spiral arms are said to trail when the ridge line of the spiral lags
with respect to the rotation direction as the radius increases, for
which $\alpha>0$ is conventional.  Assessing whether spiral arms lead
or trail requires both kinematic data on which side of the minor axis
is approaching and which receding and a determination of which side of
the galaxy's major axis in projection is physically tilted towards the
observer, which is generally determined through visual inspection of
dust lanes.  Historically, \citet{Slip22} used his first Doppler shift
measurements of galaxy rotation to conclude that all the spirals
trailed in his small sample of galaxies.  \citet{deV58} found trailing
arms in all 17 galaxies for which he had complete data.
\citet{Pasha85} found just two cases of leading spirals among the 109
galaxies he examined, and both were tidally disturbed.  While a
handful of other galaxies with leading arms have been found
\citep[\eg][]{Buta92}, there have been no significant recent updates,
and it is widely assumed that trailing spirals are the norm.

\subsection{Amplitude Estimates of Spiral Arms}
\label{sec.obs_amp}
Images of spiral galaxies in both NIR and MIR wavebands have been used
to estimate the mass contrast between the arm and inter-arm regions.
For example, \citet{RR93} measured the contrast for M51, the
prototypical grand design spiral, to be a factor of 2-3.
\citet{ElEl11} used S4G images to survey arm contrasts across a wide
range of spirals, finding a similar range (0.3-1.3 magnitudes, or
factors of 2-3), noting that grand design spirals have larger
contrasts than do flocculent spirals, and that the mean contrast
increases slightly toward later Hubble types.  \citet{Quere15}
corrected these 3.6-$\mu$m images for dust emission, which reduced
arm-interarm contrasts by some 10\% \citep{Bittner17}.  Although dust
corrections are important, a concern with all these measurements is
that emission from young stars also remains bright in the MIR, causing
surface brightness differences to overestimate surface density
variations.  This concern was addressed by \citet{Zibetti09}, who
modeled the stellar population pixel by pixel to estimate the stellar
mass surface density in a sample of nine galaxies from The Spitzer
Infrared Nearby Galaxies Survey \citep[SINGS][]{Kenn03}. The lower
$M/L$ in the spiral arms found by \citet{Zibetti09} reduced the
arm-interarm mass contrast from that in any single photometric band;
specifically they found the arm-interarm mass contrast in NGC 4321 was
half that in either $i$- or H-band images.

If spiral arms represent significant mass over-densities, they should
should also cause detectable deviations from smooth circular motion of
stars and gas.  High spatial and velocity resolution data are required
to measure such non-circular flows and only a few cases have been
reported so far \citep[\eg][]{Viss78, Kranz03, Shetty07, Erroz15}.
The data reveal ``wiggles'' in the projected isovelocity contours
across spiral arms; this is evidence that the arms are massive enough
to perturb the circular gas flow by $\sim 20\;$km~s$^{-1}$.
Spiral-driven streaming motions are best interpreted by fitting models
(see \S\ref{sec.tests}).  With the current explosion of kinematic
measurements of galaxies from integral field spectrograph surveys, as
well as gas imaging data, this may be an area ripe for more systematic
study.

\subsection{Do Spirals Trigger or Concentrate Star Formation?}
\label{sec.obs-SF}
The presence of spiral arms is observed to correlate with an
enhancement in star-formation rate (SFR), and spirals may trigger
and/or concentrate star formation \citep[SF;][]{Roberts69, KKO20}.  It
has long been noted that the average SF properties of galaxies vary
systematically along the classic Hubble spiral sequence
\citep{Kennicutt98}, with later spirals having more gas and enhanced
SF relative to earlier spirals.  \citet{Hart2017a} in agreement with
early work \citep[\eg][]{ElEl82} also note a link, with flocculent
spirals tending to be bluer than grand design spirals, but conclude
the overall SFRs are similar.  The SF efficiency (SFE) can be used
to distinguish enhanced SF due to increased gas density from
an enhancement of the SFR relative to density.  Studies of the SFE of
discs suggest it does not vary much from arm to inter-arm
\citep{Foyle10}, although \citet{YHW21} demonstrate a correlation of
SFE (or gas depletion timescales) with spiral arm strength in a large
sample.

Given the observed relationship between spirals and star formation, it
remains surprising that significant numbers of red, or quiescent
(sometimes anemic) spirals exist. The first mention of this effect was
by \citet{vdB76} who found examples of gas stripped spirals in the
Virgo and Coma clusters.  Using {\it Galaxy Zoo}, \citet{Masters10}
showed that optically red galaxies include significant fractions of
massive and/or early-type spirals, and even 6\% of late-type spirals
are red.  In general these galaxies have some residual star-formation
(\eg\ they are detectable in UV; \eg\ \citealt{F-M16}), but it is
significantly less than that expected for typical spirals of the same
size, demonstrating that spiral arms can be visible, at least for a
while, in the absence of significant star-formation
(\cf\ \S\ref{sec.cooling}).

\subsection{Spirals in High Redshift Galaxies}
\label{sec.obs-hiz}
The search for spiral patterns in high redshift galaxies is of
interest to learn when disk galaxies first became settled enough to
develop them.  The exquisite resolution of the Hubble Space Telecope
(HST) enabled observations of galaxy morphology beyond the very local
Universe, and a number of large area surveys have provided data for
understanding the galaxy population as a whole.  One complication with
high redshift observations is band shifting; high redshift images are
often in bluer rest-frame bands in which local galaxies also tend to
look clumpier (\S\ref{sec.obs-gen}).  Also genuine spirals need to be
distinguished from ``bridges and tails'' \citep{TT72} created by tidal
interactions between galaxies that are more common at higher redshift.

Early results painted a picture of high-redshift star-forming galaxies
being significantly more irregular and clumpy (\eg\ \citealt{Ab96})
and having larger velocity dispersions \citep[\eg][]{Genzel08} than
local discs. \citet{ElEl09} searched for typical spirals in a sample
of 200 galaxies out to $z\sim1.4$ in two HST surveys, finding examples
of all types of local spirals (grand design, mixed and flocculent)
alongside the more typical high redshift clumpy galaxies. Going to
even higher redshifts, \citet{ElEl14} perform a visual classification
of galaxies in the ultra-deep field, finding examples of grand design
spirals out to at least $z=1.8$, and flocculent types to $z=1.4$. They
measured the arm-inter-arm contrast in a high redshift grand design
spiral, finding it comparable to that in local spirals.
Crowd-sourcing has also been used visually to classify the largest
high-$z$ samples from HST \citep{Willett17,Simmons2017} providing
hundreds of examples of spiral galaxies at high redshift.  With the
imminent launch {\it James Webb Space Telescope}, we can expect higher
resolution and more sensitive images taken at longer wavelengths that
may reveal settled disks manifesting spirals at even higher redshift.

\subsection{The Spirals of the Milky Way}
Observing the spiral structure of our own Milky Way presents a
particular challenge because of our location within the disk.
Progress has been made via radio and IR observations, as well as
through the kinematic and astrometric observations of stars,
especially the exquisite data from the {\it Gaia} mission.  As this
topic has recently been reviewed by \citet{Vallee18} and by
\citet{ShZh20}, we will give just a short summary here.  Note that
while observations of the disk structure of the Milky Way are
challenging, they uniquely provide the full 6D phase space information
of individual stars, offering the best hope for constraining the
formation mechanism of spirals, at least in the one case of the Milky
Way.\begin{marginnote} \entry{{\it Gaia}}{An astrometric satellite
    that is returning high precision positions, proper motions, and
    photometry of billions of stars and radial velocities of the
    brighter ones}
\end{marginnote}

Surveys of gas emission lines from the Galactic plane, both of the
21cm line of neutral hydrogen and of various molecular lines, provide
intensity and kinematic information along the line of sight.
Extracting information from such data about the locations and
streaming flows in spiral arms challenging \cite[see][for a clear
  example]{Peek21}.  It is is best done by constructing models;
\eg\ \citet{Yuan69} fitted a single global 2-armed spiral while
\citet{Li16} fitted a bisymmetric spiral and separate bar flow to the
inner Galaxy only.

\citet{Reid2019} used very long baseline interferometry observations
to map the positions and motions of young, high-mass stars that appear
to be maser sources.  Their data favor a four arm model for the Milky
Way, with average pitch angles for the major parts of the arms of
$\alpha = 10^\circ$, which is surprising as four-armed spiral galaxies
are rare (\eg\ \citealt{Hart2016} find just 335 of their 6683 spirals
have four arms).  By contrast, only two major arms were revealed in
the {\it Spitzer Space Telescope's} GLIMPSE (Galactic Legacy Infrared
Mid-Plane Survey Extraordinaire) survey \citep{Churchwell2009}, which
was based on NIR star counts of the old stellar population within the
disk and may be more representative of the mass distribution.

\citet{Khop20} find surface density variations in the local star
distribution from the {\it Gaia} second data release
\citep[DR2][]{Gaia2} stellar positions, but incompleteness among the
faintest stars and a limited volume make it hard to say much about
either the mass amplitude or the global pattern.  However, the {\it
  Gaia} DR2 data also manifested ridges (or ripples) in the $R-v_\phi$
distribution of stars that \citet{Eile20} interpreted as the kinematic
signature of spiral arms; they fitted a steady spiral model to these
data to estimate the arm relative amplitude as $\sim 10\%$ and pitch
angle to be $\sim 12^\circ$.  \citet{CastroGinard21} use star clusters
identified in {\it Gaia} EDR3 to map Galactic spiral arms, their
pattern speeds and to look for age gradients.  They find a declining
pattern speed with radius, and a lack of age gradients downstream from
the arms.  Both \citet{Hunt18} and \citet{STCCR} used the phase space
distribution of stars from {\it Gaia} DR2 to test theories for the
origin spirals, which we discuss more fully in
\S\ref{sec.tests}.\footnote{The ``snail shell in phase space''
  \citep[][abstract]{Antoja18}, one of the most interesting
  discoveries in the {\it Gaia} DR2 data, is probably caused by
  excitation of a bending disturbance, rather than spiral activity.}

This somewhat confusing picture of the spiral structure of our Galaxy
unfortunately complicates interpretation of spiral arm signatures in
this one case where we have a truly close-up view.

\subsection{Summary of Observational Evidence}
\label{sec:obssummary}
This short review of the observational evidence reveals a picture
where sample selection and details of measurements
complicate any general conclusions. And while observations of the
Milky Way are revealing exquisite detail, they are from just a single
spiral galaxy.  There is clearly plenty of observational work still to
be done to improve our understanding of spiral arms in galaxies. For
example questions which at present have limited, or conflicting
results include the following:
\begin{itemize}
\item What is the distribution of pitch angles across the galaxy
  population?
\item What galaxy properties does the pitch angle physically correlate
  with?
\item What is the range of arm-interarm stellar mass contrast?
\item How large are the typical velocity perturbations caused by
  spiral arms?
\item Is SFE constant between arms and inter-arm regions in all types
  of spirals - i.e. do spiral arms trigger, or just enhance SF via
  increased gas density?
\item Does the disk of the Galaxy outside the bar have 2- or 4-fold
  rotational symmetry?
\end{itemize}
The ultimate, though still elusive, goal is to find observational
evidence that can discriminate among theories for the origin of
spiral arms.

\section{SPIRALS AS DRIVEN RESPONSES}
\label{sec.driven}
\subsection{Bar-driven Spirals}
Many spiral galaxies possess bars \citep[\eg][]{Buta15}, and both
theorists \citep[\eg][]{To69, FL73} and observers \citep[\eg][]{KN79}
have suggested bars as a driving mechanism for spirals.  A bar
introduces a quadrupole component to the gravitational field of a
galaxy that can drive an open spiral response in a smooth, massless
gas layer \citep[\eg][and dozens of subsequent papers]{SH76}, and
perhaps also a weak response in the stars \citep{Atha12}.  A good case
can be made for bar driving in the fraction of barred galaxies that
also possess (pseudo-)rings \citep{ButCom96}.  Open spirals are more
common in the majority of barred galaxies that lack outer rings, where
we should expect any bar-driven spirals to be both bisymmetric and to
have the same pattern speed as the bar.  However, spirals in the outer
disks of both simulations \citep[\eg][]{SeSp88, LCM21} and observed
barred galaxies frequently have a different pattern speed from that of
the bar, and therefore cannot be simple driven responses.  Since the
quadrupole field decays rapidly with distance from the bar, it is
likely that these spirals are excited by other means than by the bar.

Some nearby barred galaxies, such as NGC 1300 and NGC 1365, do have
beautifully regular bi-symmetric spirals joined to the ends of the
bar.  While such cases may superficially support the idea that the
spirals are driven by the bar, \citet{SpRo16} found that the spirals
in NGC 1365 have a lower pattern speed than that of the bar, thereby
ruling out the idea of simple bar driving in that case.  The
appearance of the arms starting from the bar end is not just a
coincidence, however, since \citet{SeSp88} reported that an apparent
connection between the spiral and bar lasts for a very large fraction
of the beat period.  \citet{Li16} also fitted a pattern speed for the
spiral that was lower than that of the bar in their detailed fit to
the inner Milky Way.  \citet{Font14} present estimates of corotation
radii in a large sample of galaxies based on sign changes of the
radial gas flow (see \S\ref{sec.Kaln73}), identifying multiple pattern
speeds in 28 of the 32 barred galaxies in their sample.  Furthermore,
some barred galaxies have a three-armed pattern in the outer disk,
which is inconsistent with bar driving; examples from NIR images are
M83 \citep{Ja03} and NGC~2336 (available in NED).\footnote{NASA/IPAC
  Extragalactic Database, funded by the National Aeronautics and Space
  Administration and operated by the California Institute of
  Technology.}

A number of papers report statistical evidence for or against the idea
that spirals can be driven by bars, but the conclusions are mixed.  In
studies of MIR images, \citet{Salo10}, in a reversal of the previous
conclusion by several of the same authors \citep{Buta09}, argued that
the data favor bar driven spirals, while \citet{KKC11} found that
spirals in the outer disks of barred galaxies are little different
from those in apparently unbarred galaxies. In {\it Galaxy Zoo},
\citet{Masters19} reported that arms in barred galaxies appear to be
less tightly wound on average, but a more detailed study \citep{Li21}
ruled out any statistically significant correlation, confirming the
finding of \citet{KKC11}.

In summary, spirals in some barred galaxies may be driven responses.
There are, however, many spirals in barred galaxies for which simple
bar forcing is clearly ruled out, and some other mechanism is required
to excite them.

\subsection{Tidally-driven Spirals}
A spiral pattern may also be triggered by the tidal field of a passing
companion galaxy: M51 and M81 are particularly clear examples.  The
vigor of the spiral response is generally enhanced by
swing-amplification \citep[][and \S\ref{sec.swamp}]{To81}.
\citet{SL00} and \citet{Dobb10} report detailed models of the M51
system that provide a reasonable match to most of the observational
data.  Simulations have also shown that tidal encounters can trigger
bar formation \citep[\eg][and references therein]{PeLo19}, but it is
unclear, as of this writing, what ranges of masses or orbits of
perturbers would excite spirals but not provoke bars.

\citet{KKC11} selected a sample of 13 galaxies from the SINGS survey
for which they were able to characterize the spiral pattern as either
grand design or having no well defined spiral.  They found that ``the
presence of a close companion'' defined objectively ``is (almost) a
sufficient condition'' for grand design spirals \citep[][p562]{KKC11},
confirming the earlier conclusion of \citet{KN79} from optical images.
They also note that some galaxies that lack companions (according to
their criteria) also have grand design spirals, which must therefore
be excited by other means.

Companions that may have excited spirals need not all be visible; dark
halos hosting few if any stars could also be responsible.
Hierarchical galaxy formation \citep[see][for a review]{SD15} indeed
predicts that galaxies are assembled from fragments that fall
together, and that every halo contains sub-halos having a range of
masses and orbits about the main host.  \citet{Sawa17} studied the
survival of subhalos as disk galaxies form, reporting a relative
underdensity near the center and most remaining subhalos that approach
the disk have predominantly radial orbits.  Very low-mass sub-halos
will have little effect on the disk, whereas massive sub-halos moving
on plunging orbits will disrupt the disk.  While some spiral patterns
probably are responses to a sub-halo passing the disk, as suggested
for example by \citet{Purc11}, to argue that the majority are tidally
excited transient responses would require {\it repeated} passages by
subhalos in the appropriate mass range, while the same galaxies have
so far avoided encounters with slightly more massive subhalos that
would be disruptive or trigger bars.  These requirements seem hard to
arrange, since the mass function of surviving subhalos is a smooth
power law \citep{Sawa17}.  It is more natural to suppose another
mechanism excites the majority of spirals.

\subsection{Self-excited Spirals}
Spirals are ubiquitous in disk galaxies having even a small fraction
of gas, many of which appear to lack bars or companions.  While some
patterns clearly are tidal responses, and a few may be bar-driven, we
conclude that spirals in many disk galaxies must be self-excited.
Furthermore, there can be no doubt that spirals in simulations are
also self-excited, since it is easy for the experimenter to simulate
completely isolated galaxies that lack bars, and yet such models still
spontaneously develop spiral patterns.  Simulations therefore offer a
fruitful avenue for identifying the mechanism(s) for self-excitation.

\section{SPIRAL DYNAMICS}
\label{sec.selfex}
Once the idea that spirals were gravitationally-driven density waves
in the stellar disks of galaxies took hold ({\it circa} 1964), the first
step towards an understanding of the mechanism was to examine the
gravitational stability of an axisymmetric stellar disk supported
largely by rotation.  The working hypothesis was that smooth disks
would possess spiral-shaped linear instabilities, which would give
rise to the patterns we observe.  We review aspects of spiral dynamics
in this section and discuss current theories for the origin of spirals
in \S\ref{sec.theories}.

\begin{table}[h]
\tabcolsep7.5pt
\caption{Symbols used in this review}
\label{tab1}
\begin{center}
\begin{tabular}{@{}l|l@{}}
\hline
Symbol & Meaning\\
\hline
$\alpha$ & pitch angle of a spiral \\
$\Phi$ & gravitational potential \\
$\Omega_c$ & local angular frequency of circular motion in the disk plane \\
$\kappa$ & local epicycle frequency \citep[eq.~3.30 of][]{BT08} \\
$E$, $L_z$ & specific energy and angular momentum of a star \\
$J_R$, $J_\phi$ & radial and azimuthal actions \\
$w_R$, $w_\phi$ & instantaneous phase angles of a star conjugate to the actions \\
$\Omega_R$, $\Omega_\phi$ & generalized frequencies of orbits of arbitrary eccentricity \\
$m$ & sectoral harmonic used in azimuthal Fourier analysis \\
$\lambda$, $k$ & wavelength and wavenumber of density waves \\
$\lambda_{\rm crit}$ & characteristic scale of gravitationally-driven
disturbances in disks \\
$\Sigma$ & undisturbed surface mass density in the disk \\
$f_d$ & fraction of a full-mass disk mass that is active \\
$Q$ & a numerical indicator of local axisymmetric stability \\
$\sigma_R$, $\sigma_\phi$, $\sigma_z$ & components of the stellar velocity dispersion tensor in the disk \\
$\Omega_p$ & angular frequency of a rotating disturbance, aka pattern speed \\
$\omega$ & angular frequency of a wave $= m\Omega_p (+ i\beta$) \\
$\beta$ & growth rate of an instability \\
$\Gamma$ & dimensionless shear rate \\
$X$ & dimensionless azimuthal wavelength \\
$N$ & number of particles in a simulation \\
\hline
\end{tabular}
\end{center}
\end{table}

\subsection{Preliminaries}
Here we list the principal simplifying assumptions that underlie most
theoretical analyses.  {\bf Table \ref{tab1}} gives a glossary of the
mathematical symbols used in this review.

\subsubsection{A 2D Stellar Disk in a Rigid Halo}
\label{sec.assmptns}
To this day, all analyses have assumed a razor-thin disk as a
necessary simplifying approximation.  Allowing 3D motion would not
only add an extra dimension, but would require perturbation analysis
of a 3-integral equilibrium distribution function (DF), when the third
integral itself requires numerical evaluation \citep[\eg][]{Binn16}.
Neglect of vertical motion in a thin, heavy disk may be justified by
the high vertical oscillation frequency of stars; that part of their
motion should be adiabatically invariant, and therefore decoupled from
low-frequency disturbances in the plane.  Also the in-plane part of
the behavior in simulations that allow 3D motion generally resembles
that in others in which motion is confined to a plane
\citep[\cf][]{SC84, SC14}.

Stars moving in a flat axisymmetric disk have two classical integrals
of motion: the specific energy $E$ and specific angular momentum,
$L_z$.  We will also occasionally make use of action-angle variables
$(J_R,J_\phi,w_R,w_\phi)$ because they enable an exact description of
orbits of arbitrary eccentricity.  Actions in a 2D axisymmetric
potential have a very simple physical interpretation: the azimuthal
action $J_\phi$ is identical to $L_z$, while the radial action, $J_R$,
also has the dimensions of angular momentum and quantifies the degree
of non-circular motion of a star; $J_R = 0$ for circular orbits and
increases with the orbit eccentricity.  See \citet[][their \S4]{LBK}
for a clear and concise introduction to actions, angles, and
frequencies, $(\Omega_R,\Omega_\phi) \equiv (\dot w_R,\dot w_\phi)$,
for motion in a plane. \begin{marginnote} \entry{Action-angle
    variables}{Actions are an alternative set of integrals, angles
    specify the azimuthal and radial phases of a star}
\end{marginnote}

The deceptively simple equations that govern the dynamics of a smooth
stellar fluid are the collisionless Boltzmann equation (CBE) and
Poisson's equation only; note that a collisionless fluid has no
equation of state that relates pressure to density. Since we will be
interested principally in the disk components of galaxies, we will
consider the DF of disk stars only, while the total potential is $\Phi
= \Phi_{\rm disk} + \Phi_{\rm halo} + \Phi_{\rm gas}$, where
$\Phi_{\rm halo}$ arises from the bulge and halo components, and
$\Phi_{\rm gas}$ arises from the gaseous component, which does not
obey the CBE.  To make progress, theorists have generally ignored
$\Phi_{\rm gas}$, effectively lumping it together with $\Phi_{\rm
  disk}$, which may be valid in galaxies having a low gas mass
fraction, and treated $\Phi_{\rm halo}$ as an axisymmetric, fixed
external field, which assumes that the bulge and halo are decoupled
from spiral dynamics in the disk.  This last assumption was shown to
be adequate for spiral instabilities only recently \citep{Sell21}, but
does not hold for bar instabilities (see \S\ref{sec.barstab}).

\subsubsection{Disturbance Potential}
The principal challenge is presented by Poisson's equation, for which
there are few known solutions outside of spherical symmetry, whereas
we require the potential of general non-axisymmetric density
variations in a thin disk.  The rotational invariance of Poisson's
equation allows the field of each sectoral harmonic, $m$, of the mass
distribution to be computed independently, but the radial part has no
similar useful property.  Sectoral harmonics of density and potential
remain separate at any amplitude, but the motions of the stars in
response to large potential variations generally create density
variations of several harmonics; therefore, analyses that are confined
to a single harmonic implicitly assume a disturbance of small
amplitude.

A global solution for the potential of non-axisymmetric density
variations in a razor-thin disk can be obtained by expanding the
surface density distribution in some basis set of orthogonal
functions, each of which has an exact solution for the potential, as
pioneered by \citet{Kaln71} and \citet{CB72}.

The WKB approximation treats a general wave-like disturbance as an
infinite plane wave in a razor-thin sheet. If the wave-vector lies in
the $x$-direction, the disturbed density amplitude, $\Sigma_a$, gives
rise to the disturbed potential $\Phi_1$
\begin{equation}
    \Phi_1(x,z) = - {2\pi G\Sigma_a \over |k|}e^{ikx - |kz|},
\label{eq.WKB}
\end{equation}\citep[][their equation 5.161]{BT08}.\begin{marginnote}
\entry{WKB approximation}{Invoked in quantum mechanics by Wentzel,
  Kramers, and Brillouin}
\end{marginnote}This may be applicable to spiral density waves in thin
disks if curvature of the spiral can be neglected.  Formally, this
would require the crest-to-crest wavelength, $\lambda$, to be short
compared with the distance to the center, $R$, so that $|kR|\gg1$,
with the wavenumber $k\equiv 2\pi/\lambda$, but it ``works remarkably
well'' (Tremaine, private communication) as long as $|kR| \gtsim 1$.
Note this density-potential relation holds for any angle of the wave
to the radius vector, and yields a surprisingly good approximation to
the local gravitational field near the center of a limited wave packet
because contributions to the field from the missing distant parts of
the assumed infinite wave are oscillatory and would have largely
cancelled.

\subsection{Local Stability Analysis}
\label{sec.locstab}
Dynamical studies of spiral waves in a part of the disk are described
as ``local''.  They include axisymmetric and tightly-wound spirals, as
well as open patterns.

\subsubsection{Axisymmetric Stability}
The WKB density-potential relation (\ref{eq.WKB}) was invoked by
\citet{To64} in his classic study of gravitationally-driven
disturbances in razor-thin stellar disks.  He showed that rotation
stabilizes axisymmetric disturbances in a disk lacking any random
motion unless the local radial wavelength
\begin{equation}
\lambda < \lambda_{\rm crit} \equiv {4\pi^2G\Sigma \over \kappa^2}.
\label{eq.ystick}
\end{equation}
Short-wavelength Jeans instabilities are stabilized by random motion,
and Toomre found that all axisymmetric disturbances would be stable
provided the rms radial velocity, 
\begin{equation}
\sigma_R \geq \sigma_{R, \rm crit} \simeq {3.358G\Sigma \over \kappa}
\qquad\hbox{or}\qquad Q \equiv {\sigma_R \over \sigma_{R, \rm crit}} \geq 1.
\label{eq.scrit}
\end{equation}

Equation~(\ref{eq.scrit}) applies to razor thin stellar disks.  The
constant 3.358, which results from assuming an exact Gaussian velocity
distribution among the stars, is replaced by $\pi$ and $\sigma_R$ by
the sound speed in the equivalent stability criterion for
gravitationally-driven, axisymmetric disturbances in a thin, rotating
gas sheet.  A number of authors \citep[\eg][]{BeRo88, Rome92, Rafi01}
have proposed modifications that take account of finite disk
thickness, in which the gravitational disturbance forces are weaker,
and/or a sheet containing both stars and gas.  We present an
example of instability in a two-component disk in \S\ref{sec.floccs}.

\subsubsection{Dispersion Relations and Tightly-wrapped Spirals}
\citet{Kaln65} derived a dispersion relation for axisymmetric
waves in a 2D stellar disk that may be rewritten as
\begin{equation}
    \omega^2 = \kappa^2 - 2\pi G\Sigma|k|{\cal F}.
    \label{eq.axiDR}
\end{equation}
This relation states that the frequency of the disturbance, $\omega$,
is decreased from the unforced epicycle frequency, $\kappa$, by the
self-gravity of the wave.\begin{marginnote} \entry{Dispersion
    relation}{A relation between wave number $k$ and frequency
    $\omega$ for self-consistent waves}
\end{marginnote}The ``reduction factor'' ${\cal F} \leq 1$
\citep[given by][their Appendix K]{BT08} depends upon $Q$, $k$, and
$\omega$, and quantifies the extent to which the self-gravity term is
weakened by random motion. Note, eq.~(\ref{eq.axiDR}) contains the
same essential dynamics as the study by \citet{To64}: in particular,
${\cal F} = 1$ for a cold disk ($Q=0$), giving the stability condition
on $k$ that is equivalent to eq.~(\ref{eq.ystick}).  Also ${\cal F}$
remains small enough that $\omega^2 \geq 0$ for all $k$ when $Q \geq
1$.

A similar relation was derived independently by \citet{LS66}, but in
order to relate it to spiral waves they also equated the wave
frequency, $\omega$, to the Doppler-shifted frequency at which stars
encounter an $m$-fold symmetric spiral $\omega = m(\Omega_p -
\Omega_c)$.  Here, $\Omega_p$ is the pattern speed, and $\Omega_c$ is
the circular angular frequency, which varies with radius, and the
factor $m$ appears because a star encounters $m$ wavecrests in a full
turn relative to the pattern.  By equating this forcing frequency to
the frequency of axisymmetric waves in the disk, \citet{LS66} made the
additional assumption that the wave-vector of the spiral is closely
radial, which is known as the {\bf tight-winding approximation}.
Henceforth, we will denote the WKB dispersion relation for tightly
wrapped waves,
\begin{equation}
[m(\Omega_p - \Omega_c)]^2 = \kappa^2 - 2\pi G\Sigma|k|{\cal F},
\label{eq.LSDR}
\end{equation}
as the Lin-Shu dispersion relation, or LSDR for short.

\begin{figure}
\includegraphics[width=.7\hsize,angle=0]{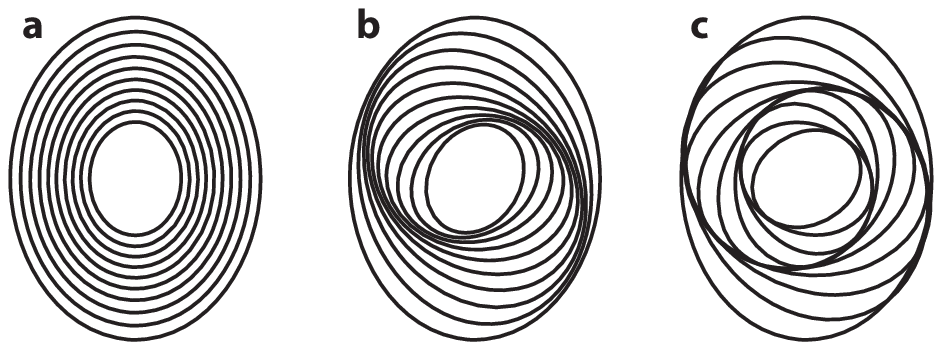}
% /home/sellwood/papers/ARAA/Kaln73.s
\caption{A collection of orbits each drawn in a frame that rotates at
  its own rate $\Omega_c - \kappa/2$, so that all close to make a
  bi-symmetric ellipse.  The left panel shows the major axes all
  aligned, whereas the ellipses are rotated by successive amounts in the
  other panels.  After \citet{Kaln73}, with permission.}
\label{fig.Kaln73}
\end{figure}

\citet{Kaln73} argued that one can think of a bisymmetric spiral
density wave as being composed of closed orbits, as shown in {\bf
  Figure~\ref{fig.Kaln73}}, each of which precesses at its angular
rate $\Omega_c - \kappa/2$.  However, the unforced angular precession
rate varies with radius and the initially aligned orbits in the
left-hand panel would wind over time, albeit at a rate that is much
slower than the shear rate in the disk.  The achievement of
\citet{LS66} was to show, within the limitations of their
approximations, that self-gravity could be used to adjust the
precession rates to create a pattern of a particular pitch angle, or
wavenumber $k$, given by eq.~(\ref{eq.LSDR}), that would not shear.
Unfortunately, neither those authors, nor anyone subsequently, has
been able explain how such a {\em steady} pattern could be established
and maintained.

Additionally, the tight-winding approximation excludes swing
amplification, which is a vital piece of spiral dynamics.  This
phenomenon was first revealed by \citet{GLB} and \citet{JT66}, who
pioneered a proper treatment of open spirals in a local approximation
(see \S\ref{sec.swamp}).  The LSDR implies a ``forbidden region''
\citep[][their \S6.2.5]{BT08} around the CR that cannot support steady
density waves for any $Q>1$, but swing-amplified waves in fact have
peak amplitude precisely where the LSDR predicts steady waves should
be evanescent.\begin{marginnote} \entry{CR}{Corotation resonance where
    $\Omega_p = \Omega_c$}
\end{marginnote}

Furthermore, eq.~(\ref{eq.LSDR}) holds equally for both leading and
trailing spirals, and therefore provides no explanation for the
preference for trailing spirals, which is both observed
(\S\ref{sec.obs=PA}) and required for outward angular momentum
transport.  Swing amplification also provides the reason that trailing
spirals are preferred.

\citet[][their \S6.2.2]{BT08} present a detailed discussion of the
LSDR despite its limited applicability to spirals in galaxies.  Its
predictions for short waves usefully yield some qualitative
indications of spiral behavior, but the fundamentally different
character of LSDR waves when $\lambda \gtsim 0.5 \lambda_{\rm crit}$,
known as the ``long wave branch'' of the relation, is of little value
for galaxy disks.

\subsubsection{Non-axisymmetric Responses}
\label{sec.responses}
There is no known general stability criterion for non-axisymmetric
disturbances in rotationally supported stellar disks, and very few
models have been shown to be globally stable.  However, before
describing global modes we first introduce two closely-related aspects
of non-axisymmetric responses to perturbations in otherwise stable
disk models: wakes and swing amplification.

\begin{figure}
\includegraphics[width=.9\hsize,angle=0]{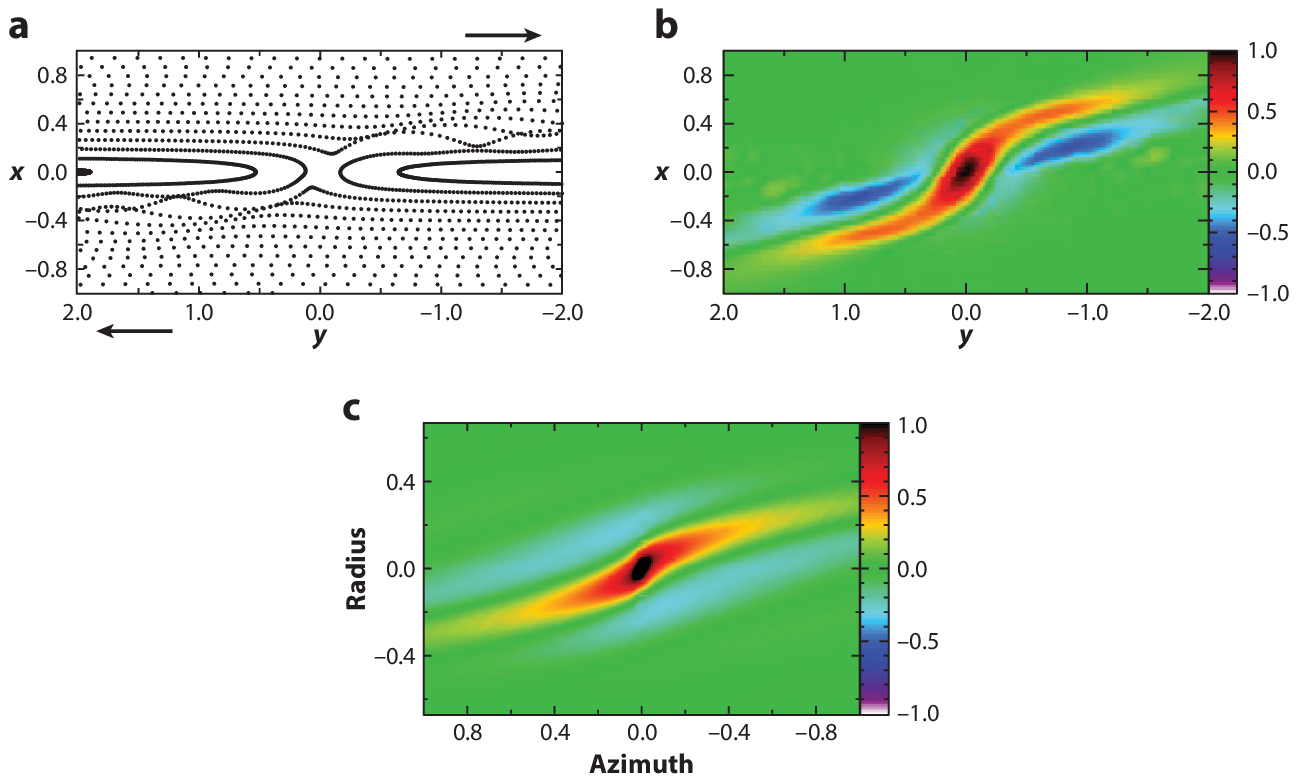}
\caption{(a) Orbits in the sheared sheet that flow past a
  co-orbiting, softened point mass ($\epsilon=0.2$) at the coordinate
  origin.  (b) The smoothed net disturbance density of the massless
  orbits in the left panel.  (c) The net response when disk
  self-gravity is included in a disk having sufficient random motion
  such that $Q=1.4$.  The unit of length in panels (a) and (b) is
  $GM/V_0^2$, where $M$ is the perturbing mass and $V_0$ the circular
  speed, whereas in panel (c) it is $\lambda_{\rm crit}$
  (eq.~\ref{eq.ystick}). Colors are relative to the maximum
  overdensity.}
\label{fig.wake}
\end{figure}

\paragraph{Wakes}
\label{sec.wakes}
The disk surrounding a co-orbiting density excess develops a trailing
spiral response \citep{JT66, Binn20}. Since both these papers are
highly mathematical, it is easy to lose sight of the physics of why
this happens, which we therefore illustrate in {\bf
  Figure~\ref{fig.wake}}.

Both papers consider disturbances in the ``sheared sheet''
\citep{Hill} and adopt a flat rotation curve model.  The approximation
focuses on a rectangular patch of the disk whose center orbits at the
local circular speed and is sufficiently small, relative to the
distance to the disk center, that the curvature of the rectangle can
be neglected.\begin{marginnote}
\entry{Sheared sheet}{An approximation first invoked by G W Hill
  for a Kepler potential}
\end{marginnote}The $x$-direction is radial and the $y$- azimuthal,
while disk material moves in a steady shear flow to the right as $x$
increases and to the left for negative $x$.  
{\bf Figure~\ref{fig.wake}a} shows how the flow is disturbed by the
gravitational attraction of a co-orbiting mass, which remains fixed at
the origin of these coordinates.  The dotted lines mark the positions
of massless particles at equal time intervals that enter the frame on
circular (\ie\ straight) orbits but are deflected as they pass the
mass.  The particles that pass at a distance experience mild impulses
that create epicyclic motion, whose effect is both diminished and
shifted farther downstream for faster moving orbits (well-spaced dots)
having larger impact parameters.  However, particles whose impact
parameters lie within the ``Hill radius'' \citep[described in][their
  chapter 8]{BT08} follow horseshoe orbits that cause them to cross
the corotation radius and reverse their apparent motion in this moving
frame.
\begin{marginnote}
\entry{Hill radius}{Region in which the gravitational field is
  dominated by the perturbing mass}
\end{marginnote}

{\bf Figure~\ref{fig.wake}b} presents the smoothed combined density of
six times as many orbits each sampled 10 times more often than those
illustrated in the left panel.  The twisted ridge of the net response
density of these massless particles results purely from their
superposed orbits in the disturbed flow.  The deflections scale with
the perturber mass, which therefore sets the spatial scale of the
upper panels.

{\bf Figure~\ref{fig.wake}c} includes the self-gravity of the disk
response as calculated by the method of \citet{JT66}, which adds
substantially to the mass of the wake, and in this case the spatial
scale is in units of $\lambda_{\rm crit}$ (eq.~\ref{eq.ystick}).  The
similarity in appearance between the response of the cold massless
disk in the top right panel, and that in the heavy, warm ($Q=1.4$)
disk reveals that the wake is induced by the gravitational deflections
of the stars as they pass the mass, augmented by the disk response.
In this case, the spatial extent of the wake is determined by the
self-gravity of the response, while the density scale varies in
proportion to the perturbing mass.

{\bf Figure~\ref{fig.wake}} is drawn for a flat rotation curve.  The
spiral response is more open where the rotation curve rises and less
open where it falls.

\begin{figure}
\includegraphics[width=.75\hsize,angle=0]{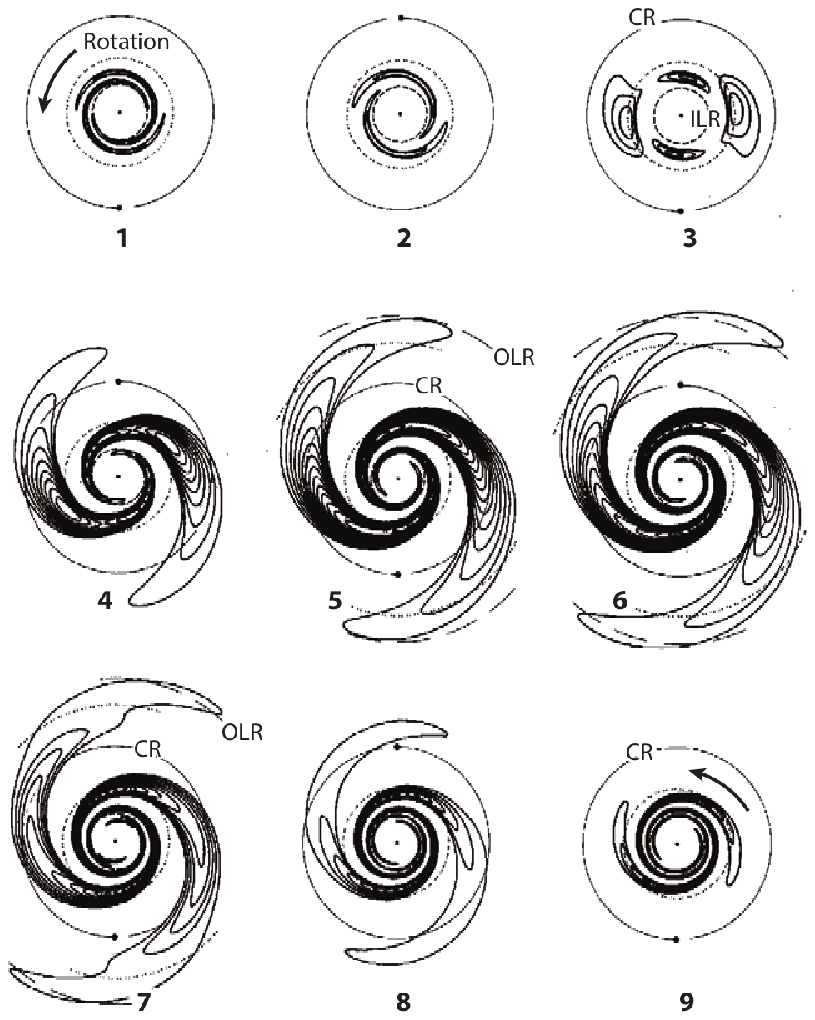}
\caption{Swing amplification in the Mestel disk.  Numbers indicate the
  time sequence in units of half a rotation period.  The initially
  imposed leading spiral unwinds at first and amplifies dramatically
  as it swings from leading to trailing.  The contours are of
  fractional overdensity only, underdense regions are not contoured.
  Abbreviations: CR, corotation resonance; ILR, inner Lindblad
  resonance; OLR, outer Lindblad resonance.  Figure adapted from
  \citet{To81}, with permission.}
\label{fig.dust-ashes}
\end{figure}

\paragraph{Swing amplification}
\label{sec.swamp}
The closely related phenomenon of swing amplification was discovered
independently by \citet{GLB} for a gaseous disk with self-gravity, the
year before the stellar dynamical treatment of \citet{JT66}.  Both
papers present a local treatment in the sheared sheet, but {\bf
  Figure~\ref{fig.dust-ashes}}, which is reproduced from \citet{To81},
gives a vivid illustration of the process in a global calculation.

This Figure results from a linearized, global perturbation analysis of
a $Q=1.5$ Mestel disk in which the surface density is reduced by $f_d
= 0.5$ so that only half the central attraction comes from the disk,
while a rigid halo makes up the other half.\begin{marginnote}
  \entry{Mestel disk}{A disk having the surface density $\Sigma(R) =
    V_0^2 /(2\pi GR)$, giving rise to a circular orbit speed, $V_0$,
    that is constant from $0\leq R \leq \infty$}
\end{marginnote}The first panel illustrates an imposed leading, 2-arm
spiral wave packet, and the subsequent panels show its evolution at
intervals of half a rotation period at the corotation radius, $R_{\rm
  CR}$, marked by the dotted circle.\footnote{Since the pitch angle of
  the disturbance changes with time, it does not have the same pattern
  speed at every radius and $R_{\rm CR}$ is therefore deduced from an
  average pattern speed.} \citet{To81} offers a long, insightful
explanation for the vigorous amplification that requires some tenacity
to follow.  We describe other aspects of the behavior seen in this
Figure in \S\ref{sec.etc}.

The factor by which an initially far leading disturbance is amplified
depends on three dimensionless parameters: $Q$, $X$, and $\Gamma$. The
familiar $Q$ was defined above (eq.~\ref{eq.scrit}), while the other
two are:
\begin{itemize}
\item $\Gamma \equiv -(R/\Omega_c)d\Omega_c/dR$. The reasonable range
  for galaxies $0 \leq \Gamma < 1.5$, where $\Gamma=0$ for uniform
  rotation and $\Gamma=1.5$ for a Kepler potential.  Note also that
  $\Gamma=1$ for a flat rotation curve.
\item $X \equiv \lambda_\phi / \lambda_{\rm crit}$, with $\lambda_\phi
  = 2\pi R_{\rm CR}/m$.  As before, $m$ is the rotational symmetry of
  the pattern and the gravitational yardstick, $\lambda_{\rm crit}$,
  was defined in eq.~(\ref{eq.ystick}).
\end{itemize}
In a disk with $\Gamma=1$ and $Q=1.2$, the amplification factor may
vary from less than 2 for $X>3$ to greater than 100 for $1 < X < 2$.
Because $\lambda_{\rm crit}$ varies with the disk surface density, $m
= 2$ disturbances are vigorously amplified in heavy disks and feebly
so in strongly sub-maximum disks where $X \gtsim 3$. However, vigorous
swing-amplification can still occur in a sub-maximum disk for higher
values of $m$.  Taking the self-similar Mestel disk as a simple
example, we find $X=2/(mf_d)$, which implies the rotational symmetry
of a strongly swing amplified spiral is $1/f_d \ltsim m \ltsim 2/f_d$.
Quite generally, swing-amplified spiral patterns should be more
multi-armed when the disk mass fraction is low \citep{SC84, ABP,
  Hart2018}.

The amplification factor varies even more strongly with $Q$ since, at
fixed $X=1.5$ and $\Gamma=1$, it exceeds 100 for $Q=1.2$, while it is
less than 10 for $Q=2$.  Thus perturbations of any wavelength in disks
having $Q \gtsim 2$ produce only mild responses.

While \citet{To81} chose to evaluate the expected amplification in the
important case of a flat rotation curve ($\Gamma=1$), we note that the
range of $X$ for vigorous amplification is increased in declining
rotation curves, and reduced when the rotation curve rises. Vigorous
amplification occurs for $1.5 \ltsim X \ltsim 4$ when $\Gamma=1.5$
(i.e. Keplerian) and the amplitude peaks at more strongly trailing
angles, whereas for $\Gamma=0.5$, the preferred range is $0.5 \ltsim X
\ltsim 1.5$ and the spirals are more open.  Naturally, the range of
$X$ for which amplification can occur shrinks to zero in a uniformly
rotating disk ($\Gamma=0$), since disturbances are not sheared.

\paragraph{Connection Between Swing Amplification and Wakes}
As already noted, the physics of wake formation is intimately
connected with swing-amplification, and indeed the formulations of
both \citet{JT66} and of \citet{Binn20} calculate the disk response to
a co-orbiting perturber as the superposition of a continuous stream of
shearing waves.  The source of the waves is the perturbing mass; a
point mass in 2D can be represented by a uniform spectrum of plane
waves of all possible pitch angles.  The leading components of this
spectrum introduce forcing terms into the shear flow, creating leading
disturbances that amplify as they swing to trailing.  Since the
spectrum is continuous, the superposed responses create a steady
trailing wake, as was illustrated in {\bf Figure~\ref{fig.wake}}.

Note that swing amplification, illustrated in {\bf
  Figure~\ref{fig.dust-ashes}}, is computed for the $m=2$ sectoral
harmonic only, whereas the wake response to a co-orbiting perturber is
summed over all possible azimuthal wavelengths $0<X<\infty$, each of
which produces a steady response.  Clearly, the response is dominated
by wavelengths that are most strongly swing-amplified, \ie\ $1<X<2$.
While the wake response is indeed caused by swing-amplification, we
will try to reserve those words to describe features whose pitch angle
evolves, as in {\bf Figure~\ref{fig.dust-ashes}}, and to use the phrase
``supporting response'' to describe steady or growing features in the
surrounding disk, as in {\bf Figure~\ref{fig.wake}}).  The vigor of the
supporting response varies with the parameters $X$, $Q$, and $\Gamma$
in exactly the same manner as for swing-amplification.

\subsubsection{More Disk Dynamics}
\label{sec.etc}
The take home message from {\bf Figure~\ref{fig.dust-ashes}} is the
phenomenon of swing amplification, but it also illustrates several
other important aspects of spiral dynamics that will factor into our
discussions of spiral modes (\S\ref{sec.modes}) and theories
(\S\ref{sec.theories}).

\paragraph{Group velocity}
\label{sec.vgroup}
Not only does the initial spiral in {\bf Figure~\ref{fig.dust-ashes}}
change its pitch angle and amplitude over time, but the wave packet
inside CR travels outwards when leading and, later, inwards when
trailing.  Recall that the group velocity of a wave packet is $v_g
\equiv \partial \omega / \partial k$, which may be calculated from a
dispersion relation.  Using the LSDR, \citet{To69} showed that a short
wavelength packet propagates radially towards corotation when the wave
is leading and away from corotation when it is trailing.  For
completeness, the sign of the group velocity on the long wave branch
of the LSDR is predicted to be the opposite for all cases of
leading/trailing and inside/outside corotation from those on the short
wave branch.  However, as noted at the end of Section 4.2.2, the LSDR
provides a poor representation of long waves in heavy disks.

Employing the local apparatus of \citet{JT66}, \citet{To69} also
demonstrated the radial propagation of an impulsively excited wave
packet.  His numerical solutions confirmed the prediction from the
LSDR when the wave was tightly wrapped but, when open, part of the
disturbance propagated across corotation to the outer disk, as also
occurred in his later global calculation ({\bf
  Figure~\ref{fig.dust-ashes}}).

\paragraph{Lindblad resonance damping}
\label{sec.wavedamp}
As the wave packet in {\bf Figure~\ref{fig.dust-ashes}} travels inward
at late times, it becomes ever more tightly wrapped and is eventually
absorbed.  Stars at any radius in the disk experience forcing by a
spiral disturbance but, except near resonances, their orbits vary
adiabatically as a small-amplitude wave packet passes over them,
leaving no lasting change.  For near circular orbits, a Lindblad
resonance arises when the forcing frequency $\omega \equiv m(\Omega_p
- \Omega_c) = \pm \kappa$.\begin{marginnote}
\entry{ILR}{Inner Lindblad resonance}
\entry{OLR}{Outer Lindblad resonance}
\end{marginnote}The negative sign is for the ILR, where stars overtake
the wave, and the positive is for the OLR where the wave overtakes the
stars, at the local epicycle frequency in both cases.  Action-angle
variables (\S\ref{sec.assmptns}) describe orbits of arbitrary
eccentricity, for which the resonance condition becomes $m(\Omega_p -
\Omega_\phi) = l\Omega_R$, with $l=\pm1,0$ for OLR, ILR and CR
respectively.

A star in Lindblad resonance may either gain or lose random energy,
depending on both its previous epicycle size and the phase difference
between the star and the wave.  \citet{LBK} showed that, to second
order, the distribution of resonant stars gains random energy on
average at Lindblad resonances causing the wave to be damped
\citep{Mark74}.  However, there are the following two caveats:

\begin{itemize}
\item The second order increase in random motion, though tiny for weak
  disturbances, does cause a lasting change to the phase-space density
  of disk stars, creating a scratch in the DF that turns out to be
  important (see \S\ref{sec.noise}).

\item Perturbation theory predicts resonance damping of small
  amplitude waves, but larger amplitude waves cause stars to become
  trapped in the resonance (see \S\ref{sec.cavity}).
\end{itemize}

Stars near the CR move slowly relative to the pattern, and may
therefore gain or lose angular momentum, depending on their phase
relative to the potential maximum.  But they neither change their
random energy (see \S\ref{sec.Jacobi} below), nor damp the wave.

\paragraph{Angular momentum transport}
\label{sec.Lztrans}
Formally, wave action density is carried at the group velocity, but
\citet[][privately assisted by Kalnajs]{To69} showed it to be
equivalent to angular momentum.  The wave packet inside corotation in
{\bf Figure~\ref{fig.dust-ashes}} has a positive (outward) group
velocity when it is leading and a negative group velocity when
trailing.  The part of the disturbance outside corotation is less
clear from the figure, but the group velocity there is outward in the
later trailing evolution.  As a trailing wave carries angular momentum
outward, it may seem paradoxical that the group velocity inside
corotation is inward.

However, spiral disturbances, such as that in {\bf
  Figure~\ref{fig.dust-ashes}}, cannot have any net angular momentum
in a disk when no external torque is applied.  Thus as the disturbance
develops, it reduces the angular momentum of the inner disk while it
increases that of the outer.  Therefore the positive group velocity
outside corotation carries positive angular momentum outwards, while
the group velocity carries a disturbance of negative angular momentum
inwards in the inner disk, and thus a trailing spiral carries angular
momentum outwards everywhere.  This explanation was given by
\citet{LBK}, but their identification of corotation as being the
radius where the sign of the angular momentum stored in the wave
changes is not always correct: edge modes for example (see
\S\ref{sec.edgegroove}) are mostly confined within the CR, and require
the sign change to lie well interior to that radius in order that the
disturbances have no net angular momentum.

Unfortunately, this is not the whole story; a Reynolds-type stress,
which \citet[][p20]{LBK} called ``lorry transport,'' is a second
radial transport mechanism.  It is of most relevance to angular
momentum transport in the opposite sense from the gravity torque on
the long wave branch of the LSDR, and is probably of less importance
for spirals in galaxies.

\begin{figure}
\includegraphics[width=.75\hsize,angle=0]{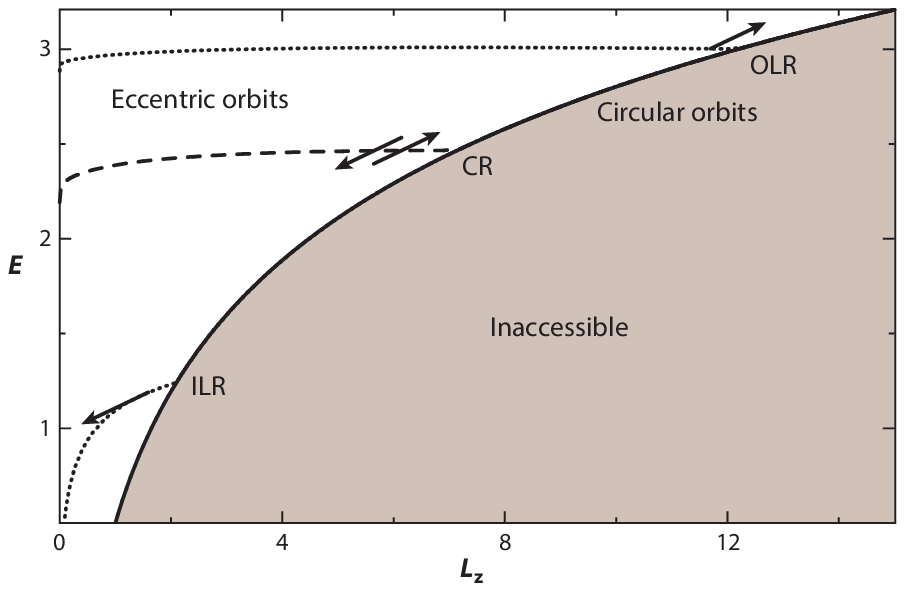}
\caption{The Lindblad diagram showing possible values of $E$ and $L_z$
  for a razor-thin Mestel disk model.  Circular orbits lie along the
  full-drawn curve and eccentric orbits fill the region above it.
  Angular momentum and energy exchanges between particles and a
  steadily rotating disturbance move them along lines of slope
  $\Omega_p$, as shown by the arrows.  The dotted and dashed lines are
  the loci of resonances, where $m(\Omega_p - \Omega_\phi) = l
  \Omega_R$, for an $m=2$ perturbation of arbitrary $\Omega_p$.}
\label{fig.Lindblad}
\end{figure}

\paragraph{Angular momentum changes at resonances}
\label{sec.Jacobi}
Stars not only gain random motion on average at the Lindblad
resonances, but they also absorb the incoming angular momentum
\citep{LBK}.  In fact, Jacobi's integral, $E_J \equiv E -
\Omega_pL_z$, which is conserved in a rotating, steady
non-axisymmetric potential \citep[\S3.3.2 of][]{BT08}, requires that
$\Delta E = \Omega_p\Delta L_z$, where $\Delta E$ and $\Delta L_z$ are
respectively the changes to the specific energy and angular momentum
of a star.  Possible changes due to one pattern all have the same
slope in the Lindblad diagram, as illustrated by the arrows in {\bf
  Figure~\ref{fig.Lindblad}}, which are directed away from the
circular orbit curve at both Lindblad resonances.  The sense of the
vectors at the Lindblad resonances in {\bf Figure~\ref{fig.Lindblad}}
illustrate the net changes, averaged over the stellar distribution
(\S\ref{sec.wavedamp}).  The locus of circular orbits is curved in the
indicated sense for any shearing model, but the curve becomes a
straight line for a uniformly rotating disk, for which no energy can
be extracted from the potential by angular momentum redistribution.

\citet{SB02} derived a useful first order relation between the angular
momentum exchanged at a resonance $\Delta L_z$ and the change of
radial action:
\begin{equation}
\Delta J_R = {l \over m}\Delta L_z \qquad\hbox{at resonances}.
\label{eq.Lchange}
\end{equation}
$\Delta J_R$ is positive at both Lindblad resonances, since the loss
of $L_z$ at the ILR, where $l=-1$, allows the star to settle deeper
into the potential well, freeing up energy for random motion to
increase.  Thus a spiral disturbance may extract angular momentum from
stars at the ILR and deposit it at the OLR, increasing $J_R$, or
heating the stars, at both resonances.

Notice that the sectoral harmonic $m$ appears in the denominator of
eq.~(\ref{eq.Lchange}), implying that a unit change $\Delta L_z$
causes less heating for disturbances of higher $m$.  This is because
the Lindblad resonances lie closer to the CR, and the outward
transport of angular momentum extracts less energy from the potential
when it is carried over a shorter radial distance.

Stars may also exchange angular momentum with the wave at CR.  Because
the vectors at this resonance are parallel to the tangent to the
circular orbit curve in {\bf Figure~\ref{fig.Lindblad}}, stars neither
gain nor lose random energy at this resonance to first order, as
setting $l=0$ in eq.~(\ref{eq.Lchange}) confirms, implying that
$\Delta E$ is exactly balanced by the change of energy associated with
the radial change to the star's guiding center caused by $\Delta
L_z$.\begin{marginnote} \entry{Guiding center}{The steadily orbiting
    point about which the star librates in its motion around an
    axisymmetric galaxy}
\end{marginnote}In a disk that is approximately uniform across CR,
the gainers and losers at that resonance roughly balance, leading to
little net angular momentum change.  However, where the density of
stars near the CR decreases steeply with $L_z$, at an outer edge say
(see \S\ref{sec.edgegroove}), the CR becomes the principal angular
momentum sink.

\paragraph{Wave action}
\label{sec.waction}
As noted above, the quantity that is transported at the group velocity
is wave action density.  The textbook example of wave action
conservation is of a wave packet travelling along a whip that has a
decreasing mass per unit length: the displacement amplitude of the
packet increases as it travels to the thin end of the whip.

As spiral waves travel radially in a disk, their amplitude is indeed
affected by the changing disk surface density, but changes to the
group velocity, which slows as the wave approaches a Lindblad
resonance, and the geometric change to the area that the wave occupies
are also important.  The focusing of an inward travelling wave causes
its relative overdensity to increase, and conversely the relative
overdensity decreases as an outward travelling wave spreads over a
larger area.  Thus the later fate of the wave packet in {\bf
  Figure~\ref{fig.dust-ashes}} appears to concentrate in the inner
disk, while the outward travelling wave beyond corotation disappears
under the lowest contour level.  This effect is particularly
pronounced for 2-arm spirals, since the Lindblad resonances, which
limit the radial extent of the pattern, lie closer to corotation for
patterns of higher rotational symmetry.  Note also that disturbance
amplitudes in the sheared sheet, such as that in {\bf
  Figure~\ref{fig.wake}}, are symmetric across corotation, since there
the disk is assumed to be uniform and curvature is neglected.

\paragraph{Super-reflection}
\label{sec.superr}
An alternative description of swing amplification is that the outgoing
leading wave super-reflects off the corotation resonance in a three
wave interaction \citep{Mark76, GT78, Drury80}.  This means that the
incident leading wave from the inner galaxy reflects as an amplified
trailing wave that propagates radially inward while conservation of
wave action requires that the reflection also excites a transmitted
trailing wave that propagates outward.  This concept is useful for our
discussions of cavity modes in \S\S\ref{sec.cavity} \& \ref{sec.waser}.

\subsection{Lumps and Scratches}
\label{sec.noise}
The collisionless Boltzmann equation embodies an idealization that
phase space is smooth; in other words, the discrete nature of stars
can be neglected and the stellar fluid is continuous.  The number of
stars in galaxy disks is large enough that this assumption holds quite
well \citep[see][for caveats]{Sell14}.  However, galaxies contain mass
clumps such as star clusters and giant molecular clouds (GMCs), and
the number of particles employed in a simulation is generally several
orders of magnitude fewer than the number of stars in a galaxy disk.
Thus both clumps in real galaxies and shot noise in simulations give
rise to inhomogeneities within the disk.

The noise spectrum in a flattened, shearing distribution of randomly
distributed gravitating masses inevitably contains leading wave
components that are strongly amplified as the shear carries them from
leading to trailing.  This behavior has two important consequences.

\subsubsection{Polarization}
The first consequence of swing-amplified shot noise is that each heavy
particle develops a trailing wake ({\bf Figure~\ref{fig.wake}}) both
towards and away from the disk center.  The wake exists in both the
background disk, and among the heavy particles themselves.  Thus the
distribution of heavy particles becomes polarized, with their
two-point correlation function being greater along the direction of
the wake and lower in other directions.  Since they are no longer
randomly distributed, the amplitude of all components of the noise
spectrum is enhanced, causing subsequent noise-induced fluctuations to
be stronger, although linear theory predicts this enhancement should
asymptote in a few epicycle periods to a mean steady excess over the
level expected from uncorrelated noise \citep{JT66, TK91}.

\subsubsection{Scratches to the DF}
The collectively amplified response to any one component of the noise
also launches a coherent wave in the disk that propagates away from
corotation \citep[][{\bf Figure~\ref{fig.dust-ashes}}]{To69} until it
reaches a Lindblad resonance where it is absorbed
(\S\ref{sec.wavedamp}).  On average, particles lose $L_z$ at the ILR
and gain at the OLR (\S\ref{sec.Jacobi}), and this outward transfer of
$L_z$ allows the wave to extract energy from the potential enabling
the scattered particles to acquire additional random energy at both
resonances ({\bf Figure~\ref{fig.Lindblad}}).  The larger amplitude
waves, in particular, therefore depopulate stars originally having
near circular orbits over the narrow region of each Lindblad
resonance, thereby creating a ``scratch'' in the DF
\citep[][p117]{SC19} and \citep{Srid19} that affects subsequent
activity.

It is important to realize that linear theory neglects this second
order effect by assumption, \ie\ it does not allow for changes to the
equilibrium state.  In fact, \citet{Sell12} found the amplitudes of
successive episodes of uncorrelated swing amplified noise in a stable
disk model rose steadily as a result of scratches to the DF.  The
Lindblad resonance absorption of each traveling wave caused an abrupt
change to the impedance of the disk at which subsequent traveling
waves were partially reflected.  Swing amplification of the weak
reflected leading wave gave a further boost to the amplitude, which
led to ever deeper scratches as the evolution proceeded \citep{SC14}.
\citet{FP15} successfully applied second order perturbation theory to
calculate this series of events, which continued until the partial
reflections became strong enough that the disk was able to support an
unstable mode \citep{Sell12, DFP19}, and coherent growth to large
amplitude began.

Here we have used the word scratch to describe quite mild changes to
the DF from weak disturbances that can cause partial reflections of
subsequent waves propagating radially within the disk.  But scattering
at a Lindblad resonance by a larger amplitude spiral could also carve
a deeper feature that seeds a groove mode (\S\ref{sec.edgegroove})
instead, and this appears to be the more usual behavior \citep{SC19}.
Even larger amplitude waves that encounter an ILR cause particles to
become trapped (\S\ref{sec.cavity}).

\subsection{Modes in Galactic Disks}
\label{sec.modes}
A normal mode of any system is a self-sustaining, sinusoidal
disturbance of fixed frequency and constant shape, save for a possible
uniform rotation; the frequency would be complex if the mode were to
grow or decay.  The perturbed surface density of a mode in a galaxy
disk is the real part of
\begin{equation}
\delta\Sigma(R,\phi,t) = A_m(R)e^{i(m\phi - \omega t)},
\label{eq.mode}
\end{equation}
where $\omega = m\Omega_p + i\beta$ is now allowed to be complex with
$\beta$ being the growth rate.  The complex function $A_m(R)$, {\em
  which is independent of time}, describes the radial variation of
amplitude and phase of the mode.

Stability analysis of a system supposes small amplitude perturbations
about the equilibrium state, which is linearized by discarding any
terms that involve products of small quantities -- see \citet{Kaln71}
for a careful formulation.  The self-consistency requirement that the
surface density variations give rise to the disturbance potential that
produced them leads to a matrix, the eigenvalues of which are the
normal modes of the disk \citep{Kaln77, Poly05, Jala07, DRV16}.  The
equilibrium is linearly unstable if any of the resulting modes have a
positive growth rate, since the disturbance exponentiates out of the
noise until the neglected second and higher order terms become no
longer negligible.

Note that the swing amplified response to a perturbation, such as in
{\bf Figure~\ref{fig.dust-ashes}}, is not a mode both because the
shape changes with time and its amplitude does not grow exponentially.
Also the wake response to an imposed co-orbiting mass clump, {\bf
  Figure~\ref{fig.wake}}, is not a mode because, to first order, it
would disperse if the clump were removed \citep[\eg][]{SC21}, and it
therefore is not self-sustaining.  Both are simply linear responses of
the disk to hypothesized imposed disturbances.  However, they are both
very helpful concepts when trying to understand the mechanisms of
self-sustaining modes.

\subsubsection{Cavity Modes}
\label{sec.cavity}
Normal modes can be standing wave oscillations that exist between two
reflecting barriers, as in organ pipes and guitar strings, which are
generally described as cavity modes in galaxy disks.  The prime
example in galaxies is the bar-forming mode, for which a reflection
takes place at the center while a super-reflection takes place at
corotation that causes exponential growth \citep[][and
  \S\ref{sec.superr}]{To81, BT08}.  Overtones also exist, but
generally have lower growth rates than the fundamental mode
\citep{To81} and are therefore less important.  Instabilities of this
type in a smooth disk are possible only if the inward traveling wave
can avoid an ILR, since linear theory \citep[][and
  \S\ref{sec.wavedamp}]{Mark74} predicts that any small-amplitude
disturbance that encounters an ILR will be absorbed, and therefore
damped.  For $m\geq2$, an ILR must be present for any reasonable
pattern speed when the center is dense, and therefore the only
small-amplitude cavity modes that are possible in a featureless disk
of this kind can only be for $m=1$ \citep{Zang76, ER98a, ER98b}.

But $\Omega_c-\kappa/2$ has a maximum value in mass models that have
gently-rising inner rotation curves, and linear bar-forming
instabilities avoid resonance damping as long as $\Omega_p > (\Omega_c
- \kappa/2)_{\rm max}$.  The dominant mode of several bar-unstable
models has been identified in simulations, with excellent quantitative
agreement of both the frequency and mode shape \citep{SA86, ES95,
  KJKJ}.  The non-linear evolution of the dominant mode is a bar in
the inner disk and a hot, mildly responsive outer disk.

\paragraph{Large-amplitude Trapping}
Note that despite the linear theory prediction that an ILR should
inhibit the bar instability, simulations having dense centers often
form bars anyway.  \citet{ELN82} emphasized this point, but a similar
result has been reported in numerous other simulations.  Many barred
galaxies are also observed to have dense bulges
\citep[\eg][]{Masters11}.  The damping of a disturbance by an ILR is a
prediction of small-amplitude perturbation theory, but a finite
amplitude disturbance at an ILR can cause particles to become trapped
in the resonance, as noted in \S\ref{sec.wavedamp}.  Swing-amplified
shot noise (see \S\ref{sec.noise}) can create sufficiently large
amplitude trailing spirals to overwhelm the ability of the ILR to damp
them.  In this case, the outcome of trapping can be a large amplitude
bar, as demonstrated by \citet{Sell89a}.  Simulations are able to
reproduce the predicted linear stability \citep[\eg][]{SE01}, but only
when they are set up carefully, employ sufficient particles that
swing-amplified shot noise can be damped, and are terminated before
the noise amplitude builds up \citep{Sell12}.

\paragraph{Stabilization by Halos}
\label{sec.barstab}
Despite years of effort, we do not understand how bars are prevented
from forming in galaxies that lack a dense center.  Historically,
\citet{OP73} argued that massive halos stabilize disks against bars,
which works because the swing amplification parameter $X>3$ for $m=2$
in sub-maximum disks, causing patterns having $m>2$ to be favored
instead (see \S\ref{sec.swamp}).  However, this strategy also inhibits
bi-symmetric spirals for the same reason, and there are a number of
galaxies, M33 being a prominent example, that have dominant 2-arm
spirals and no bar.  Indeed, \citet{SSL19} could find no satisfactory
explanation for the absence of a bar in M33, despite a systematic
exploration of many possible avenues.

The challenge presented by the apparent stability of unbarred galaxies
is further compounded because the bar-forming instabilities of a disk
in a responsive halo are more vigorous than when the disk is embedded
in an equivalent rigid halo \citep[\eg][]{Athan02, SaNa13, BeSe16}.
Since the disturbance in the disk couples to a responsive halo at
small-amplitude, the bar instability should be thought of as a mode of
the combined disk+halo system.\footnote{This first-order halo response
  differs from dynamical friction, which is second order
  \citep{BT08}.}  As such, it violates one of the assumptions of
spiral theory set out in \S\ref{sec.assmptns}.  Fortunately,
\citet{Sell21} found that a rigid halo is an adequate approximation
for groove modes, and therefore this assumption may be violated only
for bar-forming modes.\begin{marginnote} \entry{Responsive halo}{A
    halo component composed of collisionless massive particles in
    equilibrium with the disk}
\end{marginnote}

\subsubsection{Edge and Groove Modes}
\label{sec.edgegroove}
Galactic disks can also support another class of mode: edge modes
\citep{To81, PL89} and groove modes \citep{SL89, SK91} are the best
known examples, but ridges and other features are also destabilizing
(see \S\ref{sec.smooth}).  Although {\it bona fide} modes, they are
not standing waves, and instead the pattern speed is tied to the
circular orbital frequency near the radius of the feature in the
angular momentum density in the disk.

\paragraph{Edge mode} 
In the case of the edge mode, a small non-axisymmetric distortion of a
disk where the surface density decreases steeply, moves high density
material out to places where the equilibrium density is lower, and
conversely at other azimuthal phases.  On their own, such
infinitesimal co-orbiting distortions would be neutrally stable and
therefore of no interest.  But as described above (\S\ref{sec.wakes})
and illustrated in {\bf Figure~\ref{fig.wake}}, a cool surrounding
disk responds vigorously to a co-orbiting mass excess, creating a
trailing wake that extends radially far into the shear flow on either
side of the perturbing mass.  An outward bulge on the edge therefore
excites a strong supporting response from the interior disk that is
not balanced by the exterior response because the equilibrium density
drops rapidly with radius at the edge.  The forward attraction of the
interior wake on the bulging edge increases its angular momentum,
causing it to rise farther outward, and therefore to grow
exponentially as it rotates.  \citet[][pp 153-4]{To89} indicated
instability requires not only that $Q \ltsim 2$ and $X \ltsim 3$, but
also that ``the radial distance over which the disc density undergoes
most of its rapid change should be no larger than about one-quarter of
... $\lambda_{\rm crit}$'' (eq.~\ref{eq.ystick}).

In a disk with random motion, the crucial gradient is in the angular
momentum density, while random motion may spread out the surface
density gradient.  In this case, the above argument still applies to
the guiding centers, with the epicyclic librations of the stars
blurring the density variations and thereby reducing the growth rate.

\paragraph{Groove mode}
\label{sec.groove}
A groove in a disk is effectively two closely spaced edges, which
however give rise to a single mode because the distortions on each
edge are gravitationally coupled.  Again, it is the supporting
response of the surrounding disk, or equivalently swing-amplification,
that causes the groove mode to have a substantial radial extent and to
grow rapidly.  \citet{SK91} were able to obtain reasonable
quantitative agreement between their local analytic predictions and
global simulations.

\begin{figure}
\includegraphics[width=.62\hsize,angle=0]{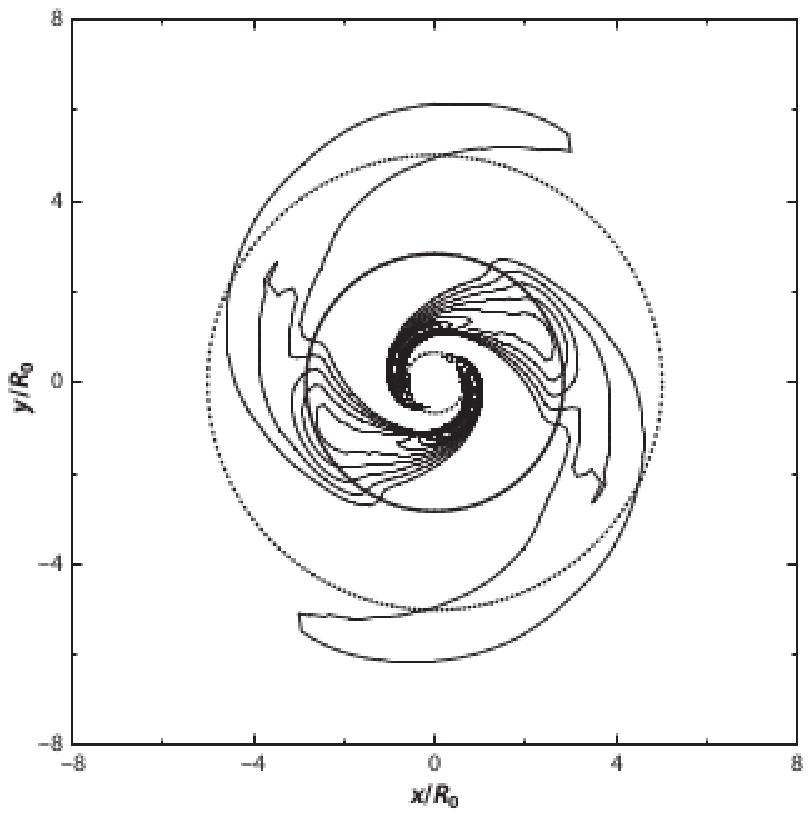}
% /home/sellwood/grooves/modefit.s 4997
\caption{The shape of the unstable mode fitted to data from simulation
  G of \citet{SC19}, in which a groove had been created by hand by
  adding random motion to particles having near circular orbits near
  $R=2.93R_0$.  The distance scales are in units of $R_0$, the central
  radius of the inner disk cutout, the solid circle marks the radius
  of CR and the dotted circles are the radii of the Lindblad resonances.}
\label{fig.gmode}
\end{figure}

A disk with a groove can still be an axisymmetric, well-mixed
(\ie\ stationary) equilibrium, but the DF is no longer a smooth
function of the integrals.  More so than for the edge mode, epicyclic
blurring can mask a groove in surface density almost entirely, since
the radial width of the groove in angular momentum is generally
smaller than the epicyclic radius of a typical disk star.
Furthermore, a deficiency need not extend to stars of large radial
action; \citet{SC19} provoked an instability in an otherwise stable
$Q=1.5$, $f_d=0.5$ Mestel disk, simply by adding random energy to
particles on near circular orbits over a narrow range centered on
$L_z=2.93V_0R_0$.  Thus their ``groove'' was merely a deficiency of
nearly circular orbits, caused by selected particles being given
additional random energy.  The shape of the resulting unstable mode,
which was determined by fitting for $A_2(R)$ in eq.~(\ref{eq.mode}) to
data from the simulation, is shown in {\bf Figure~\ref{fig.gmode}}.
Corotation for the mode, which has a fitted frequency of $\omega =
0.656 \pm 0.005 + (0.017 \pm 0.003)i$, is at $R=3.06R_0$, which is
just outside the radius of the groove they introduced to the DF.

The mode shape ({\bf Figure~\ref{fig.gmode}}) has a slight kink at the
radius of the groove.  \citet{SK91} presented mode shapes for groove
instabilities in cold disks, \ie\ lacking random motion, that had much
more pronounced kinks, which reflect the mode mechanism.  Distortions
to the two edges of the groove attract each other, and the exchange of
angular momentum causes both distortions to grow when the disturbance
on the outer edge of the groove leads that on the inner edge.  Thus,
unlike a single edge, a groove is unstable even in the absence of a
disk supporting response.  Of course, the disk supporting response
increases the mode growth rate and creates a large-scale disturbance.
The ``groove'' in a disk with random motion is a feature in the
angular momentum distribution; the mechanism is unchanged, although
random motions blur the sharp features reported by \citet{SK91} into
the mild kink visible in {\bf Figure~\ref{fig.gmode}}, and reduce the
growth rate.

\citet{SK91} also reported that CR for their groove modes lay just
outside the groove, whereas their local analysis predicted it should
lie at the groove center.  This minor difference is caused by
curvature in global modes; it decreases for modes of higher sectoral
harmonics, and disappears in the $m\rightarrow\infty$ limit of the
sheared sheet.

\section{THE ORIGIN OF SPIRALS IN GALAXIES}
\label{sec.theories}
Since spirals are ubiquitous in large disk galaxies containing a
modest gas fraction, and also develop spontaneously in simulations of
isolated disks, we argued in \S\ref{sec.driven} that some mechanism is
needed to excite them.  Unfortunately, early normal mode analyses of
apparently reasonable models of featureless disks did not identify any
promising spiral modes.  On the one hand, models where the rotation
curve rose gently from the center are dominated by vigorous
bar-forming instabilities \citep{Hohl71, Kaln78}.  On the other hand,
smooth disk models having a dense (bulge-like) center and a moderate
halo have no instabilities whatsoever \citep{To81}.\footnote{Fully
  self-gravitating disks in cusped potentials suffer from lop-sided
  instabilities \citep{Zang76, ER98a, ER98b} that cannot be blocked by
  an ILR, since $\Omega_c-\kappa<0$ everywhere.  However, they are
  inhibited by a moderate halo fraction.}  Spiral modes are still
favored today, but it took a long time to realize that the relevant
instabilities are provoked by local deficiencies in the stellar DF,
and that the earlier failure stemmed from the apparently innocuous
assumption that the DF should be a smooth function of the integrals.

As noted in \S\ref{sec.obs}, few spiral galaxies manifest highly
regular spiral patterns.  Individual arms can rarely be traced over a
significant radial range and bifurcations and branches are common.
Spiral patterns that develop spontaneously in simulations of isolated,
unbarred stellar disks generally give this impression also; thus an
understanding of the mechanism for spiral generation in the
simulations may provide a useful guide to the behavior in galaxies.

\citet{MPQ70} and \citet{Hohl71} first reported spiral patterns
appearing spontaneously in simulations of collisionless particle
disks, apparently confirming that they are a collective phenomenon of
many body Newtonian dynamics.  Subsequent simulations have progressed
from their $N \sim 10^5$ particles confined to a plane to $N \gtsim
10^8$ moving in 3D, but the qualitative spiral behavior has not
changed.  As $N$ is increased, the relative amplitude of shot noise,
which varies as $N^{-1/2}$, is reduced, enabling spiral patterns to be
traced to collective modes that stand clear of the noise
(\S\ref{sec.noise}).

\subsection{A Recurrent Cycle of Groove Modes}
\label{sec.best_theory}
It is now clear that the spontaneous development of spiral patterns in
simulations of isolated and unbarred disks results from a recurrent
cycle of groove modes \citep{SC14, SC19}.  The conceptual breakthrough
of this discovery is that it discards the assumption of a DF that is a
smooth function of the integrals, which was entrenched in all early
work.  Instead, the DF possesses a groove, or a deficiency over a
narrow range of $L_z$, that seeds a linear instability
(\S\ref{sec.groove}).  Furthermore, the nonlinear evolution of a
groove instability creates new grooves at the Lindblad resonances of
the original mode, thereby setting up a recurrent cycle.  This
behavior can occur in collisionless particle disks only and does not
have an analog in gas disks, for example.

Power spectra \citep{SA86} taken from the simulations \citep{Sell89b,
  Rosk12, Minc12, SC14, SC19} reveal that the changing appearance of
the spirals results from the superposition of several separate waves,
each having a constant pattern speed over a broad radial range.  The
amplitudes of the individual disturbances grow and decay, but each is
detectable over a period of several rotations at the corresponding
corotation radius.  These waves are modes that differ fundamentally
from those in other theories (\eg\ \S\ref{sec.waser}) because they are
supported by a vigorous disk response, they do not last for nearly as
long, and fresh instabilities develop to maintain spiral activity.

Although the individual modes have constant shapes and pattern speeds,
the spiral appearance of a simulation changes continuously.  This is
because the disk supports several modes at any one time, each having a
different pattern speed and perhaps also angular periodicity, as well
as a time varying amplitude.  The superposition of several modes
causes the pitch angle of individual arm features to decrease with
time \citep{SC21} (see {\bf Supplemental Video 1}),\footnote{
  temporary url:
  http://www.physics.rutgers.edu/$\sim$sellwood/supp\underline{~}material.html}
while fresh patterns come to the fore, and the detailed appearance of
the overall pattern changes radically in less than one disk rotation.

The recurrence mechanism, which was clearly demonstrated by
\citet{SC19}, is as follows: as a groove mode saturates, the angular
momentum stored in the wave (\S\ref{sec.Lztrans}) drains at the group
velocity (\S\ref{sec.vgroup}) onto the Lindblad resonances where it is
absorbed (\S\ref{sec.wavedamp}), scattering resonant stars to more
eccentric orbits (\S\ref{sec.Jacobi}), thereby depopulating another
part of the DF over a narrow range of $L_z$ and low $J_R$, and seeding
a fresh groove instability having a new pattern speed.  The initial
groove to seed such a cycle in a real galaxy could be caused by
resonance scattering as, say, an orbiting mass clump settles into the
disk or by the near passage of a small companion or, in the unlikely
circumstance that neither of these events happen, spiral disturbances
could bootstrap out of the noise \citep{Sell12}.

\citet{SC19} showed that scattering at both the Lindblad resonances of
any one mode created grooves in the DF that seeded fresh groove-type
instabilities, with corotation for each subsequent mode being close to
the newly-carved grooves.  Thus a new instability could be either
closer to or farther from the disk center and, moreover, need not have
the same angular symmetry as the original.  Even starting from very
contrived initial conditions that supported a single instability only,
the disk quickly developed many new instabilities that caused the
usual apparent transient spiral evolution.

A recurrent cycle of groove modes has been firmly established in
simulations, but it is not easy to find evidence that it operates in
real galaxies.  The best evidence is that the distribution of
particles in action space \citep{SC14} acquired multiple scattering
features resembling those in the \Gaia\ data from the local Milky Way.
See \S\ref{sec.tests} for this and other possible tests.

\subsubsection{Disk Heating by Spirals}
\label{sec.heating}
Note that scattering of stars at Lindblad resonances not only carves
grooves, but increases the general level of random motion in the disk,
thereby rendering the disk less responsive to future instabilities.
Thus spiral activity in a purely stellar disk is self-limiting, and
simulations of massive disks suggest it fades on a time-scale of some
ten disk rotations \citep{SC84, SC14}.  Spiral activity can persist
``indefinitely'' \citep[][p3]{SC14} if the disk is cooled, as
discussed in \S\ref{sec.cooling}.

\subsubsection{Sub-maximum disks}
A slower heating rate was reported by \citet{Fuji11} and others in
their simulations of sub-maximum disks, and those authors wrongly
blamed the more rapid heating reported by \citet{SC84} on collisional
relaxation.  \cite{SC14} dismissed that idea and explained instead
that the amount of Lindblad resonance heating, \ie\ $\Delta J_R$ for
outward transport of a given $\Delta L_z$, decreases with increasing
$m$ (eq.~\ref{eq.Lchange}).  Therefore less rapid heating is expected
in halo-dominated disks that favor more multi-arm spirals (see
\S\ref{sec.swamp}).

\subsection{Other Theories}
\label{sec.others}
A number of other theories have been proposed for the origin of spirals
in galaxies.  Here we review three of them.

\subsubsection{Quasi-steady Density Waves}
\label{sec.waser}
The ubiquity of spirals in galaxies led many astronomers
\citep[\eg][]{Oort62} to favor long-lived spiral patterns, since they
would not require constant regeneration.  This preference was met by
the widely-cited theory of quasi-steady waves promoted in the book by
\citet{BL96} and the review by \citet{Shu16}.  Following
\citet{Mark77}, these authors argued that ``grand design'' spirals in
galaxies are manifestations of a cavity-type (\ie\ WASER) mode in a
sub-maximum disk that is dynamically cool over most of the disk, but
which also possesses a ``$Q$ barrier'' \citep[][p212]{BL96} both to
provide an inner turning point and to shield the
ILR.\begin{marginnote}
\entry{WASER}{wave amplification by stimulated emission of ``radiation''}
\end{marginnote}The mildly-unstable mode persists for many tens
of galactic rotations and becomes quasi-steady due to dissipative
shocks in the gas; they also allowed that superposition of a second
mode may be needed in some cases.  As noted in \S\ref{sec.swamp},
strong swing-amplification occurs for $1 \ltsim X \ltsim 2$.  By
considering only bi-symmetric disturbances in sub-maximum disks,
\citet{BLLT} exploited the mild disk response when $X > 3$ in order to
obtain slowly-growing spiral modes in their stability analysis of many
galaxy models.  Note that their mode calculations included a global
solution for the gravitational field and invoked a fluid model for the
disk, which is valid away from Lindblad resonances.

Simulations by \citet{Sell11} of one of the cases presented by
\citet{BLLT} confirmed that a single, slowly-growing mode was present
when disturbance forces were restricted to $m=2$.  The basic state of
the collisionless particle disk did not evolve in this restricted
simulation while the mild instability grew slowly.  Not surprisingly,
however, \citet{Sell11} also found much more vigorous instabilities
appeared when higher sectoral harmonics contributed to disturbance
forces, and the contrived $Q$-profile of the disk, which was designed
to support the $m=2$ mode, was rapidly changed.  The onset of disk
heating by these multi-armed disturbances was increasingly delayed as
larger numbers of particles were employed because those instabilities
took longer to grow from the decreased shot noise.  The inclusion of
gas cooling, which his simulations omitted, would have slowed the disk
heating rate, and allowed the multi-arm activity to persist
indefinitely (see \S\ref{sec.gas}).  Therefore, the slowly growing
bi-symmetric spiral modes in halo-dominated disks determined by
\citet{BLLT} would indeed be overwhelmed by true vigorous
instabilities having $m>2$.

Furthermore, the \Gaia\ DR2 data \citep{Gaia2} revealed a rich level
of substructure in the phase space distribution of stars near the Sun,
indicating that the local disk of the Milky Way is far from the
settled, well-mixed state invoked by \citet{BL96}.  \citet{STCCR}
argued it seemed unlikely that a\Ignore{ ``delicate'' \citep[the
    adjective used by][]{BL96}} spiral instability could flourish in
the observed disequilibrium state of the Milky Way disk, and also
showed that rival theories would {\em naturally create} some of the
observed features in phase space, whereas quasi-steady modes would
not.  Thus if the Bertin-Lin mechanism for spiral generation were
somehow to operate in the Milky Way, some other recent and/or on-going
disturbances would be required to create the observed unrelaxed phase
space \citep{STCCR} without interfering with the spirals.
Consequently, the theory is now beset with multiple serious issues.

\subsubsection{Responses to Noise}
\label{sec.lumps}
\citet{To90} abandoned the idea of spirals as normal modes, and
advocated instead that a collection of massive clumps in the disk,
each of which becomes dressed with its own wake (\S\S\ref{sec.wakes}
\& \ref{sec.noise}), would create a ``kaleidoscope'' of shearing
spiral patterns.  Local simulations of this process by \citet{To90}
and \citet{TK91} employed a modest number of particles confined to a
shearing patch, in which the particles themselves were the ``massive
clumps''.  \citet{DVH} conducted global simulations of a sub-maximum
disk composed of $10^8$ star particles, embedded in a rigid halo, to
which they added a sprinkling of heavy particles.  Since responses in
their sub-maximum disk favored $6 \ltsim m \ltsim 12$
\citep[][\S\ref{sec.swamp}]{DVH}, the seed particles induced evolving
multi-arm spiral patterns in the stars.  In separate experiments they
also tried a single perturber, which they removed after its wake had
developed, and reported continued spiral activity without additional
forcing.  As linear theory predicts that a wake in a stable disk
should decay once the driving term is removed, \citet{DVH} attributed
the continuing activity to non-linear effects.

However, the perturber might have seeded unstable modes that would
have continued to grow after it was removed.  \citet{SC21} therefore
reproduced their experiment, but were unable to find any coherent
modes in the on-going activity.  Taking one step further, these
authors tried a single co-orbiting mass in the stable \citep{To81}
half-mass Mestel disk model in which responses are most vigorous for
$2 \leq m \leq 4$.  In this case, they found the disk had acquired
several discrete instabilities after the perturbing mass was removed;
this was in contrast to the behavior in the low-mass disk.
\citet{SC21} were able to show that the supporting responses at $m=2$
\& 3 carved isolated grooves in the heavier disk, but scattering at
higher $m$ resonances in the halo-dominated disk blurred together to
create a broad feature that was not destabilizing.

\citet{To90} and \citet{DVH} suggest that ``ragged'' \citep[][p473]{To77}
spirals in galaxies result from responses to co-orbiting giant
molecular clouds, massive star clusters, \etc, and to the lingering
disk responses should any disperse.  Although their numerical results
are sound, it is unlikely that their proposed mechanism accounts for
the observed spirals in galaxies for several reasons.  First, the
heaviest perturbing mass, $10^7\,M_\odot$, that \citet{DVH} employed
produced only a modest wake within a narrow annulus in their
halo-dominated disk.  Second, \citet{DVH} found that the disk response to a
collection of randomly placed heavy particles was multiple spiral
arms, not one that was predominantly 2- or 3-armed.  Third, clumps
massive and numerous enough that their associated wakes produce
large-amplitude and radially-extensive spiral patterns would scatter
disk stars causing rapid heating of the disk so that the responses
would fade quickly unless the disk were cooled (see
\S\ref{sec.cooling}) aggressively, and the necessary cooling
\citep{To90} seems rather extreme.

Spirals in real galaxies (\S\ref{sec.obs}) generally have greater
amplitude, radial extent, and lower rotational symmetry than those in
the simulations of \citet{DVH}, suggesting that more massive clumps in a
more massive disk would be needed.  But it is likely that more massive
disks readily support unstable spiral modes, as discussed above
(\S\ref{sec.best_theory}), obviating the need to stretch responses to
mass clumps into a full theory for spirals in galaxies.

\subsubsection{Shearing Spirals}
\label{sec.misguided}
A number of authors \citep[see the review by][for early
  references]{DB14} and also \citet{Kawa14, Baba15, KN16, MK18, MK20},
have argued that spiral patterns are hardly density waves at all, but
wind more tightly over time at a rate that is almost as rapid as if
they were material features.  These papers report evidence that
swing-amplification plays a prominent role in the development of the
spirals, as was first noted by \citet[][their Fig.~3]{SC84}.  In fact
{\em all} agree that the pitch angle of individual features in
simulations decreases over time, that spiral patterns change
continuously and differ beyond recognition after a single disk
rotation.

However, this apparent behavior can be manifested by the superposition
of two or more\break long-lived, uniformly-rotating patterns, as was
convincingly demonstrated by \citet{SC21}.  They presented an
animation, available as {\bf Supplemental Video 1}, showing the time
evolution of the net disturbance density when two steady wake
responses having differing pattern speeds were superposed.  All that
is required for the appearance to resemble swing amplification is that
the steady waves partially overlap in their radial extent and the peak
amplitude of the pattern having the higher pattern speed lies interior
to that of the slower.  Since the dynamical clock runs faster towards
the center, this second requirement is almost inevitably
satisfied.\footnote{See \citet{LCM21} for a counter example, in which
  the bar was slowed by dynamical friction, causing the faster
  rotating spiral in the outer disk of their models to appear to
  alternate between leading and trailing.}

We stress that power spectra extracted from simulations over a few
disk rotations almost always reveal that apparently shearing,
transient waves are decomposed into a few steadily rotating waves each
extending over a radial range centered near the CR.  We described the
origin and nature of these underlying modes above
(\S\ref{sec.best_theory}).

Note that if the only process were swing amplification, the input
noise would merely be amplified by a factor (\S\ref{sec.swamp}),
causing the resulting spiral amplitudes to scale as $N^{-1/2}$.
\citet{SC14} reported that initial spiral amplitudes indeed scaled
as $N^{-1/2}$, but the amplitudes quickly rose to a level that was
independent of the number of particles, which they varied over several
orders of magnitude.  They also noted that it took longer to reach the
common amplitude as $N$ was increased.  The most natural explanation
of their findings is that their models were linearly unstable to
spiral modes that exponentiate out of the noise, which is reduced as
$N$ is increased, and the modes saturate at a common amplitude due to
nonlinear effects.

None of the papers that claim spiral patterns in simulations to be
simply swing-amplified transients have reported the effect of varying
the number of particles by a few decades, yet the visible spirals in
their simulations have similar amplitudes even in experiments having
several million particles.  It seems highly likely that the spirals
they have reported were created by instabilities and the shearing
patterns and apparent swing amplification resulted from superposition
of some number of unstable modes.

\subsection{Observational Tests of Theoretical Ideas}
\label{sec.tests}
Observational evidence, reviewed in \S\ref{sec.obs_amp}, indicates
that most spirals are density waves.  Both NIR photometry and 2D
velocity maps clearly suggest they are quasi-sinusoidal density
variations in the underlying stellar disk that are massive enough to
create non-axisymmetric streaming flows in the gas. \citet{Shu16}
reviews multiple papers that attempt to fit spiral models to nearby
galaxies.  However, this exercise tells us nothing about the origin of
the density waves and, since gas rapidly adjusts to changes in the
spiral potential, the flow pattern is also insensitive to spiral
lifetimes.

Spiral arms have long been predicted to trigger star-formation either
via shocks to the gas clouds \citep{Roberts69}, or simply because the
gas flow converges \citep[\S\ref{sec.Kaln73} and][]{KKO20}.  When the
spiral has a fixed pattern speed over a broad radial range, the
streaming speed of stars and gas relative to the wave increases
with radial distance from corotation, and leads to shallower
stellar age gradients among the newly-formed stars downstream from the
spiral arm; this was first proposed by \citet{Dixon71} and restated by
\citet{DP10}.  While correct, we emphasize that even swing amplified
disturbances that shear at close to the material rate for a while,
develop wave-like properties in the later stages ({\bf
  Figure~\ref{fig.dust-ashes}}) when gas and stars stream through the
arms.  The contrasting prediction of quasi-steady spiral structure is
that the pattern speed is constant over the entire radial range of the
spiral.  \citet{Foyl11} rule out age gradients downstream from the
spiral in their sample, but others \citep{Chan17, YH18, Mill19,
  Pete19} claim to have detected them.  However, none of these careful
studies was able to establish a fixed pattern speed over the entire
radial extent of the spiral.

A further suggestion from the same authors \citep{PD19} is that $\cot
\alpha$, with $\alpha$ being the pitch angle of the spiral, should
have a uniform distribution across some range of $\alpha$ and over
spiral arms in many galaxies if spirals wind up over time.
\citet{Li21} applied this test to a sample of 200 galaxies, finding
$\cot \alpha$ values that were consistent with a uniform distribution
over the range $15^\circ \ltsim \alpha \ltsim 50^\circ$.

Both these tests could possibly distinguish the classic density wave
theory of a single large-scale spiral mode (see \S\ref{sec.waser})
from other models, but winding spirals are predicted in all three of
the other mechanisms discussed in \S\S\ref{sec.best_theory} \&
\ref{sec.others}.  Thus, a different kind of test is required to
discriminate among theories of how the spiral disturbances are
excited.  Currently, the only foreseeable such test can be made within
the Milky Way, and requires the exquisite data from {\it Gaia}.

As noted above, the second data release from {\it Gaia} \citep{Gaia2},
has revealed extensive substructure in the phase space distribution of
stars near the Sun.  \citet{Hunt18} showed that some of the features
in the velocity distribution could be reproduced by the winding spiral
model.  However, \citet{STCCR} converted the coordinates to
action-angle variables, finding a number of coherent features in
action space that sloped to smaller $L_z$ with increasing $J_R$, as
expected from ILR scattering.  They also found a highly non-uniform
distribution in angles, which is clear evidence that the stellar
distribution is not well-mixed, and has therefore been subjected to
recent disturbances.  However, the features in action space are
unaffected by phase mixing and should therefore endure, although
scattering by molecular clouds may gradually blur them.  These authors
experimented with idealized models of possible perturbations and
concluded that the observed features were somewhat more consistent
with transient spiral modes, than a simple model of a winding spiral.
It is also likely that some, though not all, of the features in
action-space were created by resonances with the bar of the Milky Way
\citep{Mona19}.

Thus {\it Gaia} DR2 has not provided sufficient evidence to
discriminate conclusively among the different theories for the origin
of the spirals. However, it is to be hoped that future releases with
more precise measurements extending to greater distances from the Sun
may one day afford a decisive test.

\section{GAS IN SPIRAL GALAXIES}
\label{sec.gas}
Our discussion so far has ignored the gas component (except as
possible mass clumps), even though our description of the observations
(\S\ref{sec.obs}) noted that at least a small gas fraction seemed
almost essential for isolated galaxies to possess spiral patterns.
Here we discuss three roles for gas in spiral galaxies.

\subsection{Maintaining Spiral Activity}
\label{sec.cooling}
Both stars and gas clouds are scattered by the spirals
(\S\ref{sec.heating}), but while stellar random motions cannot be
damped, those of the gas component are.  Individual clouds collide
dissipatively, with the collision energy being radiated, which drives
them toward non-intersecting streamlines that are circular in an
axisymmetric potential.

Unfortunately, simulations are unable to model gas properly because
the dynamical processes of spiral formation occur on spatial scales
that are many orders of magnitude greater than would be required to
capture the full physical behavior of the clumpy, multi-phase
interstellar medium (ISM).  The sub-grid physics of fragmentation,
star-formation, feedback, heating, cooling, shocks, turbulence,
metallicity increases, magnetic fields, \etc, can be modeled only by
adopting \adhoc\ rules.  However, as far as spiral dynamics is
concerned, more or less any rule that mimics dissipation prolongs
spiral activity \citep[\eg][]{SC84, CaFr85, To90, Rosk12, ABS16}.
Cosmological simulations of galaxy formation also mimic gas physics
and support mild spiral patterns (see \S\ref{sec.galform}).

Not only do the gas clouds themselves dissipate random energy, but new
stars are formed with the kinematics of this lower velocity dispersion
component.  Thus the crucial low velocity dispersion population of
stars is augmented, thereby maintaining the responsiveness of the star
plus gas disk and enabling spiral activity to persist.  Without
replenishment, star formation would eventually consume the gas, with
much of it being locked away in low mass stars having essentially
infinite lifetimes.  However, a drizzle of infalling gas onto the
galaxy disk, over and above any possible fountain flow resulting from
star formation activity \citep[see \eg][]{RB19}, not only replenishes
the gas, but it also gradually raises the disk surface density, which
diminishes $Q$ (eq.~\ref{eq.scrit}) and makes the disk more
responsive.  \citet{SC84} found that a rate of gas infall and star
formation of a few solar masses per Earth year over the entire disk of
a galaxy would provide sufficient cooling to balance the heating by
moderate spiral activity, which is consistent with the requirement to
maintain star formation rates, first noted by \citet{LTC80}.  Thus the
observation \citep[\eg][and \S\ref{sec.obs}]{Oort62} that almost all
spiral patterns are seen in galaxies that contain gas and are forming
stars can be understood by this argument.

Hierarchical galaxy formation \citep[reviewed by][]{SD15} indeed
predicts late infall both as cooling of shock-heated gas and in cold
flows onto the disks of galaxies in the field, which is responsible
for the inside-out growth of galaxy disks.  Galaxies in large
clusters, however, may have not only their ISM stripped by their
relative motion through the hot intra-cluster gas, but also their
disks are deprived of fresh infalling cool gas, which is at least part
of the reason that clusters host many S0 galaxies and few spirals, as
proposed by \citet{Gunn82}.  However, there are two reasons that
cluster S0s should not have the properties of field disk galaxies that
have merely been deprived of gas: first, galaxies in a cluster
originated in a denser environment than those in the field, causing
them to have generally larger classical bulges (from hierarchical
merging) and second, accretion of gas to grow the disk will have
stopped at an earlier stage, causing them to have less extensive
disks, on average.  Note that S0 galaxies also exist in the field, and
\citet{F-M18} propose two mechanisms for their origin: faded spirals
for low mass S0s, and mergers to create those of higher mass.

\subsection{Gas Flows in Spiral Potentials}
\label{sec.Kaln73}
{\bf Figure \ref{fig.Kaln73}}, in \S\ref{sec.locstab} above, was also
used by \citet{Kaln73} to illustrate gas streamlines in spirals, since
cold gas will settle onto the illustrated ballistic orbits if they can
be nested without intersecting, although a shock must intervene where
orbits cross.  From these diagrams, one can see that the flow
converges as the gas approaches the spiral and, if the gas overtakes
the wave (inside CR), it flows inwards in the arms, whereas outward
flow along the arm is expected outside corotation \citep{Kaln73}.
This sign change of the radial flow velocity within spiral arms was
exploited by \citet{Font14} to identify the radii of CRs in many
galaxies.  Note that these closed streamlines create flows in the
opposite sense between the arms, and there is no net inflow or
outflow, at least in the absence of shocks.

Although offering valuable insight, this picture is highly idealized,
and the detailed dynamics of the ISM matters a great deal.  Since
global simulations cannot begin to model local star formation,
feedback, \etc, \citet*{KKO20} adopt an intermediate course, and try
to build a more detailed picture of ISM behavior in a small part of
the disk of a spiral galaxy that is subject to an imposed spiral
perturbation.  Their still idealized model predicts that stars are
preferentially formed in spiral arms, as a result of the converging
gas flow.  They also show that supernova feedback blows holes in the
ISM, creating chimneys, and the shear flow carries the rims of the
larger holes to create features that match the observed spurs and
feathers.

\subsection{Flocculent Spirals}
\label{sec.floccs}
The theories reviewed above (\S\ref{sec.theories}) addressed the
origin of patterns having moderate numbers of spiral arms.  However,
flocculent galaxies have many short spiral arm segments
\citep[\S\ref{sec.obs-gen} and][]{Sand61}; the prototype is NGC~2841,
for which the spiral fragments stand out in blue light, while NIR
images are almost featureless \citep{Bloc96}.  NGC~5055 is also
flocculent in blue light, but some IR images reveal an underlying
2-arm spiral that was confirmed kinematically by \citet{ThLe97}.

\begin{figure}
\includegraphics[width=.7\hsize,angle=0]{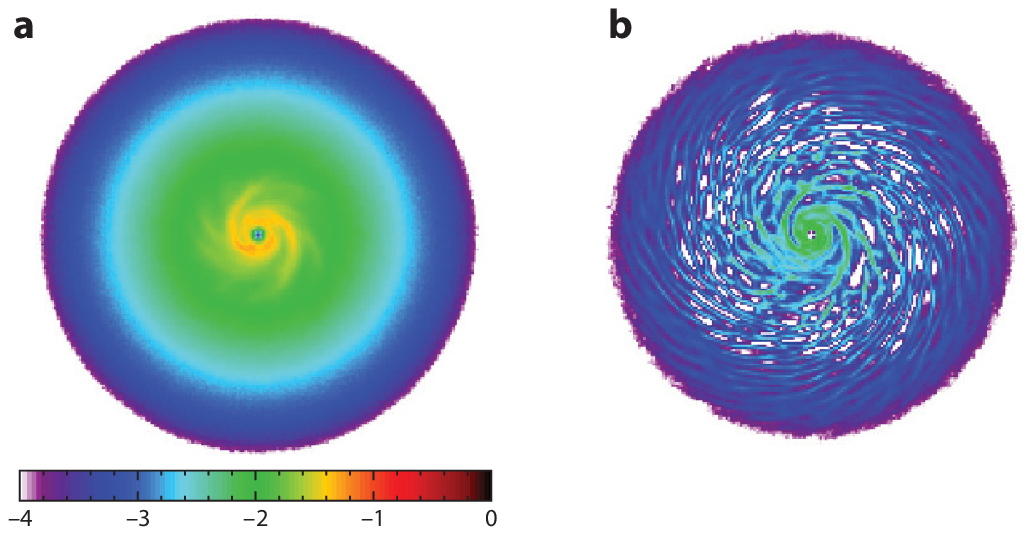}
% /home/sellwood/floccs/analys.s
\caption{A simulation having a two-component Mestel disk.  (a) shows
  the massive warm component at the same instant as the low-mass very
  cool component (b), which manifests a flocculent spiral pattern.
  The radius of the disk is $20R_0$, and the color scale reports the
  logarithm of the surface density in units of $V_0^2/(GR_0)$.}
\label{fig.floccs}
\end{figure}

\citet{EEL03} relate flocculence to turbulence in the ISM, but a more
interesting dynamical explanation was proposed earlier by
\citet{ElTh93}, who suggested that flocculent patterns arise through
gravitational instabilities in a low-mass cool disk component.  They
presented simulations of a low-mass disk embedded in a massive halo
that manifested flocculent spirals.  Here we show ({\bf
  Figure~\ref{fig.floccs}}) that a two component disk behaves in a
similar manner.  The model employed to create {\bf
  Figure~\ref{fig.floccs}} has a half-mass ($f_d=0.5$), $Q=1.5$,
Mestel disk, while the cool disk, also composed of collisionless
particles, has one-tenth the mass ($f_d=0.05$) and an initial
$Q=0.05$, which is a deliberately low value in order to mimic the
dynamical responsiveness of a gas-rich component.  The two components
are dynamically decoupled at first, and the supporting response favors
sectoral harmonics $20 \ltsim m \ltsim 40$ in the cool component.  The
cool disk quickly creates flocculent spirals, perhaps driven by mass
clumps as suggested by \citet{To90} and DVH13 (\S\ref{sec.lumps}),
while the warm disk, which would have been stable \citep{To81} in the
absence of the cool disk has a mild spiral with a few arms; some
coupling between the two components is apparent where the cool
component has heated significantly in the inner disk by the time shown
($t=100R_0/V_0$).

A two-component disk like this could perhaps arise naturally if the
old disk were starved of fresh gas for a while, and then suddenly
accreted a substantial gas component.  The flocculent instabilities
would trigger star formation in the arm segments perhaps giving rise
to a galaxy resembling NGC~2841.  A prediction of this model is that
there should be a dearth of intermediate age stars in the disk of a
flocculent galaxy.

\section{ROLE OF SPIRALS IN GALAXY EVOLUTION}
As noted at the outset of this review, spiral activity is a driver of
evolution in a disk galaxy.  Thus the present day structure of galaxy
disks is not merely the result of initial conditions at the time of their
formation, but it has been changed, at least in part, by
internally-driven evolution, as has long been argued by \citet{Korm79}
and reviewed by \citet{KK04}, although they stressed the role of bars
over spirals.

\subsection{Radial Migration}
\label{sec.rm}
For years, attention focused on Lindblad resonance scattering by
spirals, and changes at corotation went unreported.  \citet{SB02} were
therefore surprised to find that a transient spiral mode causes
greater angular momentum changes to stars at the CR than occur at the
Lindblad resonances.  The angular momentum gains and losses by
different stars at the CR generally roughly balance.

\begin{figure}[t]
\begin{center}
\includegraphics[width=.9\hsize,angle=0]{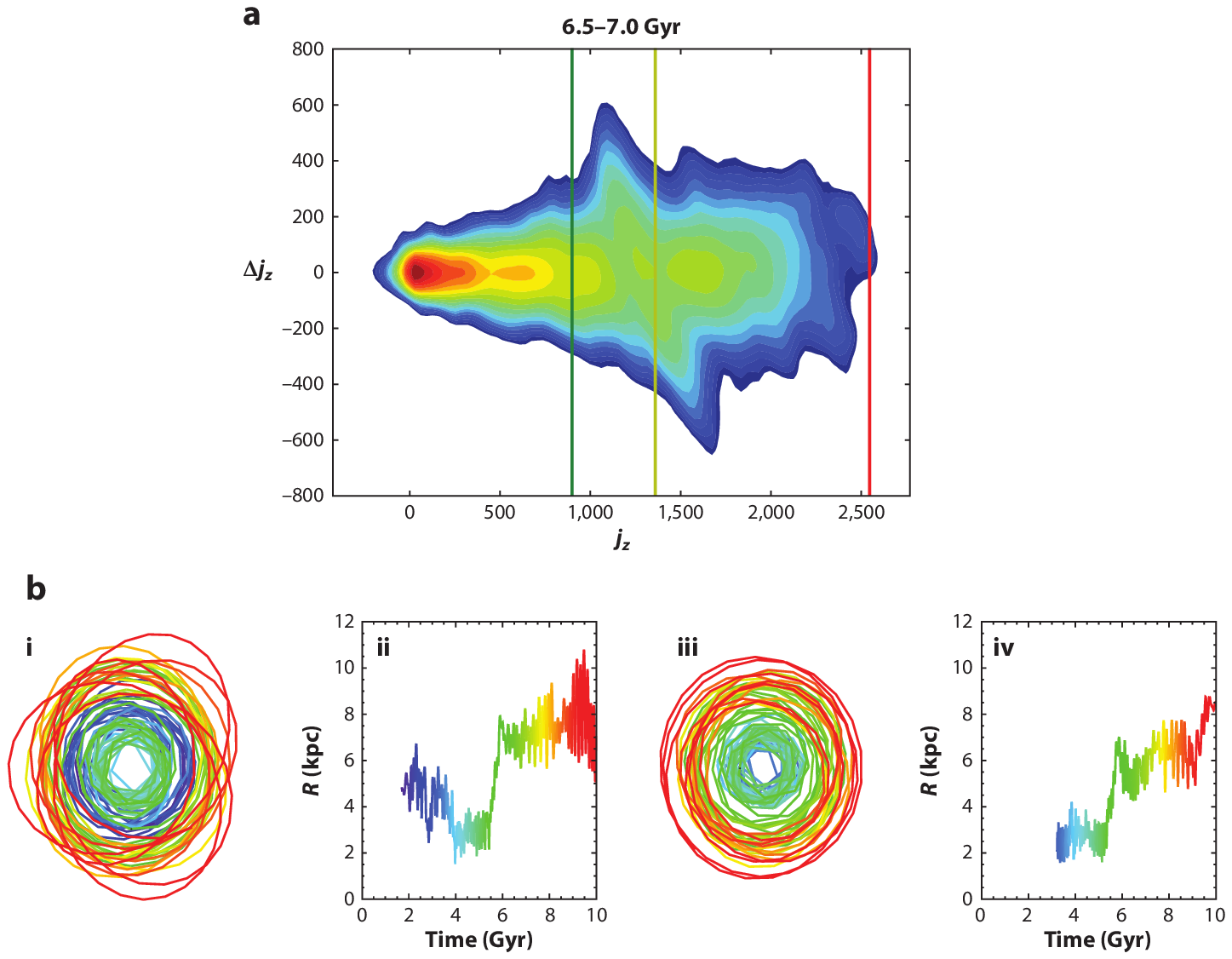}
\end{center}
\caption{(a) The distribution of changes to the angular momentum, $\Delta
  j_z$, by 0.5~Gyr later for the given $j_z$ at 6.5~Gyr.  The vertical
  lines mark the estimated $j_z$ of corotation for the dominant three
  spiral modes over the time interval 6 to 7 Gyr.  (b) Two examples of
  orbits from the same simulation.  The colors of the lines in
  subpanels {\it i} and {\it iii} change with time as indicated in
  subpanels {\it ii} and {\it iv}, which show the instantaneous radius
  of the particle.  Figure adapted from \citet{Rosk12}, with
  permission.}
\label{fig.dLz}
\end{figure}

An example from a more realistic simulation is shown in {\bf
  Figure~\ref{fig.dLz}a} \citep[adapted from][who used $j_z$ for
  $L_z$]{Rosk12}.  The distribution of changes manifests an inclined
ridge in the middle of the range of $L_z$ indicating $\Delta L_z$
values for some particles that range up to $\sim L_z/2$ at the start
of this interval.  The vertical lines mark the estimated $L_z$ values
for corotation of three separate spiral modes in the disk in their
simulation over a longer interval, 6 to 7 Gyr, but the middle wave
dominates over the time interval reported in {\bf
  Figure~\ref{fig.dLz}a} (see their Fig.~6).  It is clear that many
particles having smaller $L_z$ values than the middle vertical
(yellow) line increased their initial $L_z$ to rise outwards across
corotation, and others initially having larger $L_z$ values lost
similar amounts also to sink inside corotation.  The numbers of
gainers and losers were similar, and the slope of the ridge indicates
that there was a tendency for particles that crossed CR to end up
equally as far in $L_z$ from the resonance as before.

{\bf Figure~\ref{fig.dLz}b} presents two orbits of particles in the
same simulation that have experienced substantial radial migration.
The orbit in the left pair of panels at first migrates inwards at
$t\sim 4\;$Gyr and then outwards by a larger amount at $t\sim 6\;$Gyr,
as does the orbit in the right pair of panels.  Notice that in all
three instances, migration is rapid and occurs with no significant
increase in the size of the orbit's epicycle, indicated by the short
period oscillations.

These $L_z$ changes, which completely dwarf those at the Lindblad
resonances, had not attracted attention because they do not heat the
disk (eq.~\ref{eq.Lchange}), and stars largely change places in a
dynamically neutral manner.  However, radial mixing of stars with
different chemical abundances has important consequences for modeling
the radial distribution of elements in a galaxy disk \citep{Rosk08,
  SB09}.

\subsubsection{Mechanism of radial migration}
Stars near corotation move slowly with respect to the spiral
perturbation and therefore experience almost steady forcing from the
wave, which allows large changes to build up -- a process that is
analogous both to surfing on ocean waves and to Landau damping in
plasmas, although the consequences differ.  Stars orbiting just behind
the density excess are attracted forward by it and gain angular
momentum.  However, the result of gaining angular momentum is that the
star moves onto an orbit having a larger guiding center radius, and
its angular frequency about the center decreases.  If the star were
just inside corotation, and gaining on the density excess, the change
can cause it to rise to a radius just outside corotation where it
begins to fall behind.  This behavior is described as a horseshoe
orbit; {\bf Figure~\ref{fig.wake}a} includes a few example orbits
whose motion reverses in the rotating frame.  Conversely, stars just
ahead of the perturbation are pulled back, lose angular momentum and
sink inwards, where they orbit at higher frequency.  Those outside
corotation, where the perturbation gains on them, could lose enough
angular momentum to cross corotation and begin to run ahead of the
wave.  As long as the gradient $\partial f/\partial L_z|_{J_R}$ is
fairly shallow, roughly equal numbers of stars gain as lose, and they
largely change places.

Were the spiral potential to maintain a fixed amplitude, stars on
horseshoe orbits would be described as trapped.  As they move slowly
with respect to the wave, stars would take a long time to reach the
next density maximum, where the changes just described would be exactly
undone.  However, if the amplitude of a transient spiral mode has
decayed by the time the star reaches the next density peak, it will no
longer be trapped and will continue to move with a lasting change to
its angular momentum.

The radial extent of the region where horseshoe changes occur varies
as the cube root of the perturbation amplitude \citep[eq.~8.91
  of][]{BT08}, and therefore widens as a perturbation grows.
\citet{SB02} found the spiral was strong for less than half the
horseshoe period for most trapped stars, which consequently undergo a
single change.

The process affects stars with small peculiar velocities most
strongly, since greater epicyclic motion leads to less coherent
forcing by the spiral potential \citep[\eg][]{DaWy18}.  Also
\citet{SSS12} showed, and \citet{Kord15} found supporting empirical
evidence, that migration is only mildly reduced by vertical motion.
This is because the vertical excursions of even thick disk stars are a
small fraction of an open, low-$m$ spiral's crest-to-crest wavelength,
$\lambda_\perp = 2\pi R\sin\alpha/m$, where $\alpha$ is the pitch
angle of the spiral.  Note that the potential of a WKB wave
(eq.~\ref{eq.WKB}) decays away from the disk mid-plane as
$\exp(-2\pi|z|/\lambda_\perp)$.

\subsubsection{Random Walk in Radius}
Changes to the guiding center radii caused by a series of transient
spiral modes with corotation radii scattered over a wide swath of the
disk will cause some stars to execute a random walk in radius, while
preserving radial and vertical actions \citep[\eg][]{Mikk20}.  Typical
step sizes range to over $\sim 2\;$kpc \citep{SB02, Rosk12, ABS16},
though they are smaller, and consequently cause weaker churning
\citep{SB09}, for lower amplitude spirals, such as occur in
simulations having hotter and thicker disks (see \S\ref{sec.galform}).
The resulting churning of the stellar distribution has implications
for abundance gradients and age-metallicity relations.
\begin{marginnote}
\entry{Churning}{The shuffling of the angular momenta
  of disk stars by radial migration}
\end{marginnote}

\citet{Minc12} suggest that bars play a role in radial mixing, which
they argue is enhanced by overlap between the resonances of the bar
and spirals.  Note, however, that bars themselves tend not to be
transient disturbances, and therefore stars that are trapped in a CR
with the bar will repeatedly cross and recross that resonance.
Indeed, the trapping of stars near corotation of the bar could be
important for the maintenance of inner rings \citep{ButCom96}.  An
essential aspect of radial migration by spirals is that the pattern
has decayed before there is time for the star to make a second
crossing of the CR of a spiral, leading to a lasting change in $L_z$.
Resonance overlap could perhaps provide a route for particles trapped
in the CR of the bar to escape to the outer disk.

The underlying metallicity gradients are also {\bf blurred} by
epicyclic motions.  Since the guiding center radius of a star is
determined only by its angular momentum, the intrinsic radial gradient
without blurring can be estimated from samples of Milky Way stars
without having to integrate their orbits.
\begin{marginnote}
\entry{Blurring}{The smearing effect caused by epicyclic motions of
  the stars that may reduce an underlying radial metallicity gradient}
\end{marginnote}

\subsubsection{Radial Migration in Sub-maximum Disks}
\label{sec.LMD}
Simulations of sub-maximum disks support multi-arm patterns (as noted
in \S\ref{sec.lumps}) and therefore differ from the low-order
rotational symmetry preferred in galaxies (\S\ref{sec.prefm}).  The
crest-to-crest wavelength $\lambda_\perp = 2\pi/k_\perp$ of such
multi-arm patterns is generally much shorter than that in all but the
most tightly wrapped $m=2$ spirals, which has several consequences
that reduce radial migration.
\begin{enumerate}
  \item For a fixed density amplitude, $\Sigma_a$ in
    eq.~(\ref{eq.WKB}), the spiral potential is weaker for larger
    $k_\perp$, which will diminish the radial extent of the horseshoe
    orbit region that is responsible for migration.
  \item The period of a horseshoe orbit trapped in the CR depends on
    both its frequency difference from CR and the azimuthal distance
    between wave-crests, which is shorter for higher-$m$ spirals.
    Since migration relies on the spiral having already decayed before
    a star makes its second horseshoe turn, those stars having periods
    long-enough to make only a single turn are confined to a narrower
    region about CR, implying a smaller average step size for
    migration.
  \item The increased value of $k_\perp$ causes the spiral potential
    (eq.~\ref{eq.WKB}) to decay away from the mid-plane more rapidly,
    lessening its ability to affect the orbits of thick disk stars.
\end{enumerate}
These factors, which stem from the short wavelength of the spirals,
will reduce the extent of churning that is possible in both the thin
and thick disks in simulations of atypically sub-maximum disks, as
reported by \citet{Vera-C14}.

\subsubsection{Observational Evidence for Radial Migration}
Many papers have claimed observational evidence both for and against
radial migration but not all are equally convincing.  So far, all
evidence is indirect, although one of the science goals of GALAH
\citep[GALactic Archeology with HERMES;][]{DeSi15} and other upcoming
Galactic spectroscopy surveys is to use detailed chemical tagging to
identify stars born from the same molecular cloud, and to examine
their distribution throughout the Milky Way \citep[but see
  also][]{Casamiq21}.

Three papers stand out: \citet{Kord15}, using RAVE (Radial Velocity
Experiment) data \citep{Stei06}, found supersolar metallicity stars
having lowish eccentricity orbits in the solar neighborhood and argued
they must have migrated from the inner disk.  \citet{Hayd15}, using
Apache Point Observatory Galactic Evolution Experiment (APOGEE) data
\citep{Maje17}, measured the metallicity distribution functions (MDFs)
across a large volume of the Milky Way disk having a radial and
vertical extents of $3 < R < 15\;$kpc and $|z| < 2\;$kpc respectively.
They found a striking systematic change with radius to the shape of
the midplane MDF and concluded that radial migration was the most
likely explanation for the shape of the MDF in the outer Galaxy.
\citet{Fran20} fitted a model of churning and blurring to APOGEE red
clump stars, concluding that the secular orbit evolution of the disk
is dominated by diffusion in angular momentum, with radial heating
being an order of magnitude lower.

\subsection{Flattening Rotation Curves}
\label{sec.flatten}
The rotation curve, or circular speed as a function of radius, is
remarkably smooth for most galaxies \citep[\eg][their Fig 5 in the
  html version only]{Lelli16}.  There is barely a feature even where
the central attraction shifts from being baryon-dominated to dark
matter-dominated, which \citet[][p828]{Kent87} described as a
``disk-halo conspiracy.''  A few authors \citep[\eg][]{Kaln83, Kent86,
  PW00} have drawn attention to ``bumps and wiggles''
\citep[][p205]{Free92} in long-slit rotation curves, some of which
correspond to photometric features in the light profile.  While this
is undeniable evidence for significant mass in the disk, the
underlying cause of these small-scale features may be spiral arm
streaming rather than substantial fluctuations in the radial mass
profile of the disk.

Spiral instabilities may also be responsible for featureless rotation
curves, as first argued by \citet{LH78}.  \citet{BeSe15} conducted
experiments with growing disks in which they artificially chose to add
material having a narrow range of angular momentum. Some of their
models had a dense central mass and all had a (rigid) cored outer
halo.  They found that no matter what the angular momentum of the
added particles, the mass distribution in the disk rearranged itself
such that the resulting rotation curve became remarkably featureless.

\subsubsection{Smoothing Mechanism}
\label{sec.smooth}
\citet{BeSe15} presented a more controlled experiment in which they
added particles to the stable Mestel disk to build a narrow ridge.
The spirals that developed in this model were the result of two
unstable modes that were provoked by the density ridge.  Local
stability analysis of an axisymmetric ridge-like density excess
\citep{SK91} predicts that, for each sectoral harmonic, the normal
modes are wave pairs with corotation on opposite sides of the ridge.
The most rapidly growing pair of modes was for $m=3$ in their
simulation, which was preferred by the disk supporting response (see
\S\ref{sec.swamp}).  As the mode amplitudes rose, horseshoe orbits
(\S\ref{sec.rm}.1 and {\bf Figure~\ref{fig.wake}a}) developed at both
CRs but, unlike in a featureless disk, the presence of the ridge
caused the resulting $L_z$ changes to be strongly out of balance in
opposite senses for each separate mode; naturally, the combined effect
of both modes did not change the total $L_z$ of the disk.  Since CR
scattering removed far more particles from the ridge than were added
to it, the ridge was erased and the rotation curve was flattened
almost perfectly.

\subsubsection{Maximum Entropy State}
Thus it seems that the distribution of angular momentum in the
baryonic material that created a galaxy disk does not need to be able
to account for the featureless character of most galaxy rotation
curves.  Any small-scale variations in a reasonable distribution will
be erased by spiral activity.

\citet{HTR17} developed a maximum entropy formalism to determine the
expected surface density profile in a disk in which radial migration
efficiently scrambles the angular momenta of individual stars, while
preserving the circularity of their orbits and the total mass and
angular momentum of the disk.  They showed that the maximum entropy
surface mass profile was not perfectly exponential, but nevertheless
corresponded well with the surface brightness profiles of a large
sample of disk galaxies.  Since disk galaxies generally possess
population and/or color gradients, they cannot have fully reached the
maximum-entropy end state, but nevertheless their mass profiles appear
to be close to this idealized model.

\subsection{Driving Turbulence in the ISM}
\label{sec.magfld}
It has long been recognized \citep[\eg][]{Rees94}, that the origin of
the large-scale magnetic field in galaxies presents a challenge, in
that the standard $(\alpha,\Omega)$ dynamo mechanism \citep{Park55}
has difficulties creating large-scale fields of the observed strength
from the expected seed fields.  The process uses differential
rotation, the $\Omega$-effect, combined with turbulence in the ISM,
the $\alpha$-effect, to amplify the field.  Calculations
\citep[\eg][]{Ferr98} that invoke turbulence driven by mechanical
input to the ISM even from clustered supernovae struggle to achieve
the observed field strengths primarily because the turbulence is
driven on too small a scale.  More recent work is thoroughly reviewed
by \citet{KhKr18}.

\begin{figure}
\includegraphics[width=.9\hsize,angle=0]{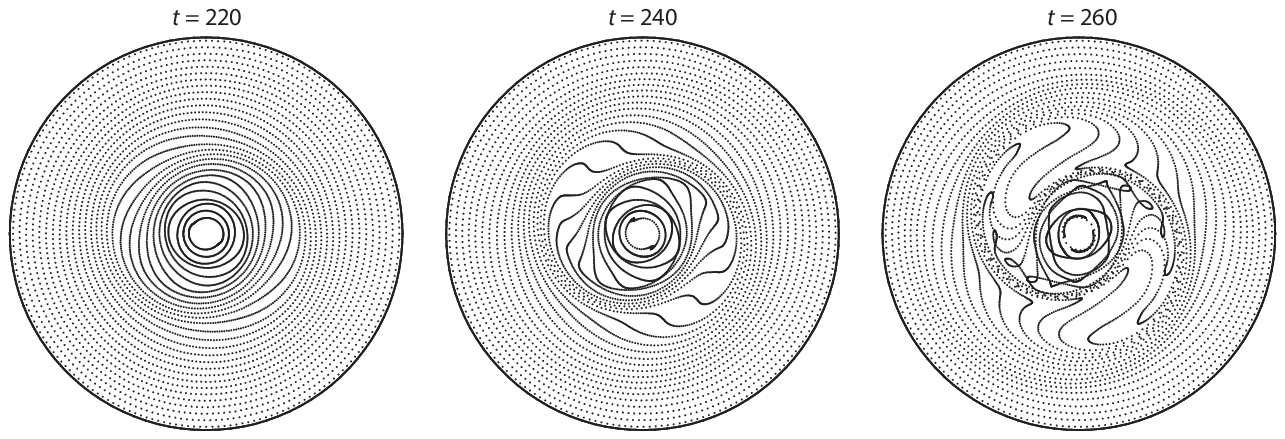}
\caption{Part of the evolution of test particles that began on rings
  having initially circular orbits in the groove-unstable model used
  by \citet{SB02}.  The rings of particles can be thought of as
  tracing streamlines up until the moment they intersect.  Reproduced
  from \citet{SePr02}.}
\label{fig.slines}
\end{figure}

However, transient spiral waves churn not only the stellar
distribution (\S\ref{sec.rm}), but also the ISM.  The three snapshots
in {\bf Figure~\ref{fig.slines}} \citep[from][]{SePr02} show part of
the evolution of rings of test particles that began on initially
circular orbits in the groove-unstable model used by \citet{SB02}.  In
the period shown, the growing instability causes distortions to the
rings that are most pronounced near the CR of the instability.
Imagining the non-interacting particle rings to trace gas streamlines,
with individual clouds threaded by magnetic field, it is clear that
the evolving spiral potential creates crossing streamlines, at which
point collisions between gas clouds would occur.  Note that particles
from widely differing radii encounter each other, and that the spiral
instability drives turbulence on a much greater radial scale than
would supernovae.  \citet{SB02} therefore suggested that the slow
magnetic field amplification from supernovae alone could be
accelerated by this source of turbulence on grander scales.

Unfortunately, this suggestion has yet to be subjected to a convincing
test.  Although spiral-driven turbulence may well have contributed to
the promising magnetic field amplification reported by \citet{Pakm17},
their simulations included too many physical processes to be able to
isolate the role of non-axisymmetric gravitational forces arising from
spiral arm evolution.  The simulations by \citet{KhKr18} included
self-gravity of the magnetized gas only, but adopted the gravitational
field of an imposed, steadily rotating spiral potential, which
crucially omits the evolving gravitational field that is important to
driving turbulence by radial migration, while \citet{WB21} employed a
sub-maximum disk that developed multi-arm spirals that are unable to
drive large radial excursions (\S\ref{sec.LMD}).

\subsection{Age-velocity Dispersion Relations}
\label{sec.avr}
\citet{Wiel77}, and others, pointed out long ago that the random
motions of older disk stars in the Milky Way are greater than those of
younger ones.  The evidence was enormously strengthened and extended
by \citet{Mack19}, who made use of the stunning improvement to stellar
kinematics from {\it Gaia} DR2, abundance analysis from APOGEE, and
state-of-the-art techniques to assign ages to stars.  Their sample,
which they separate it into high and low $[\alpha/{\rm Fe}]$,
extends over a broad radius range and vertical distance from the
mid-plane.  \begin{marginnote} \entry{$[\alpha/{\rm Fe}]$}{A
    logarithmic measure, relative to solar values, of the
    $\alpha$-element abundance relative to that of iron.}
\end{marginnote}

\begin{figure}
\includegraphics[width=.9\hsize,angle=0]{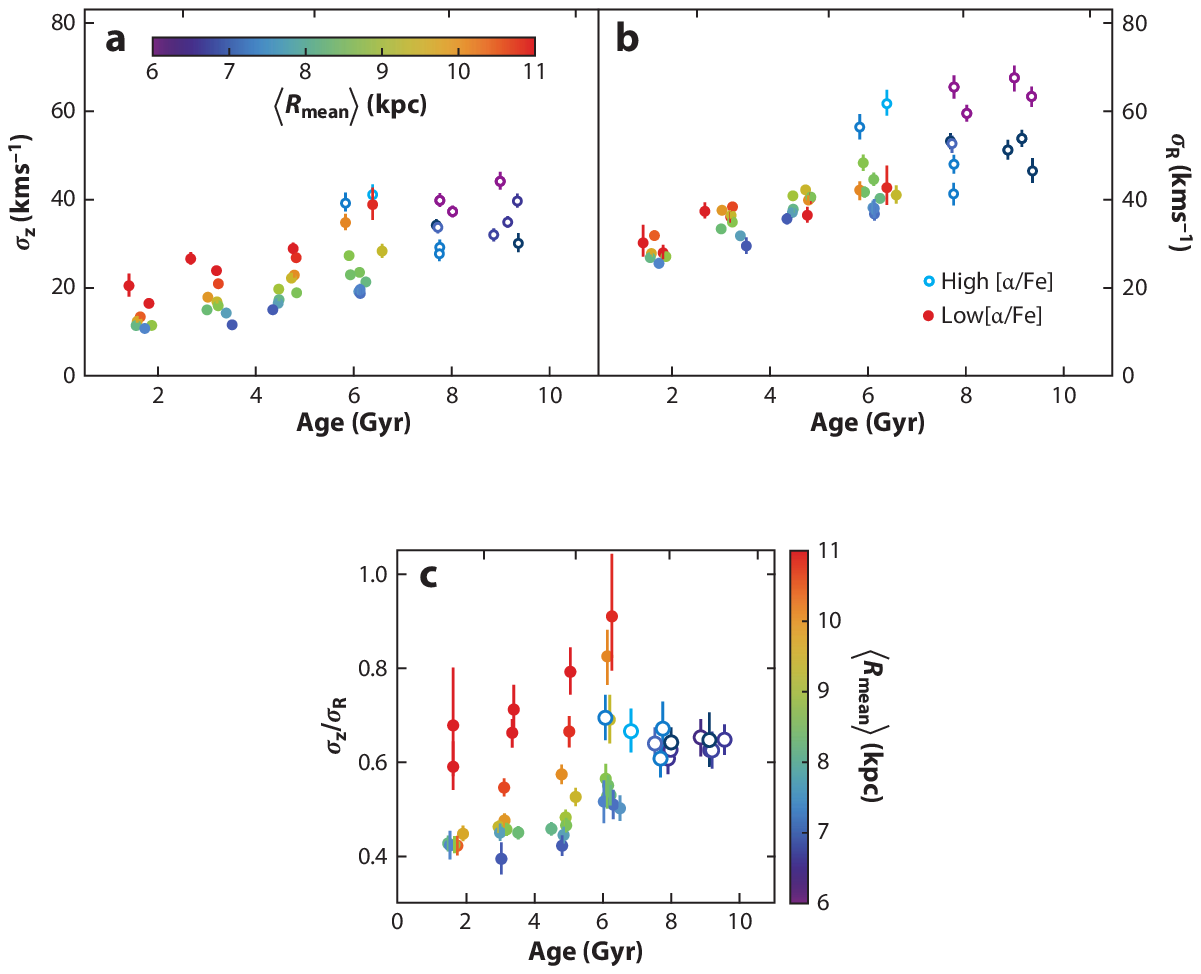}
\caption{Dispersions of the vertical (a) and radial (b) stellar
  velocities divided into age bins and color coded by Galacto-centric
  radius. The age bins for each group of points are the same but the
  points have been shifted slightly in mean age to separate them more
  clearly.  (c) The ratio of vertical to radial axes of the stellar
  velocity ellipsoid.  Figure adapted from figs 5 and 7
  of \citet{Mack19}, with permission.}
\label{fig.AVRs}
\end{figure}

Two of their principal findings are reproduced in {\bf
  Figure~\ref{fig.AVRs}a,b} clearly reveal the kinematically distinct
and older population of the high $[\alpha/{\rm Fe}]$ stars, which in
their sample are predominantly in the inner Milky Way.  Also, older
stars have larger velocity spreads at all radii, with the high
$[\alpha/{\rm Fe}]$ stars having distinctly larger
$\sigma_R$.\footnote{The third velocity dispersion component,
  $\sigma_\phi$, is strongly coupled to $\sigma_R$ through the
  epicycle motions of stars \citep{BT08} and is therefore not
  independent.}  \citet{Mack19} fitted power laws to the variations of
dispersion with age, finding an index of $\sim 0.3$ for $\sigma_R$
that is almost independent of $[\alpha/{\rm Fe}]$.  The bottom panel
reveals that the velocity ellipsoid of the low $[\alpha/{\rm Fe}]$
stars is quite flattened in the inner disk, though less strongly with
age, and becomes much rounder at larger radii.

\citet{Sell14} reviewed the mechanisms that have been invoked to
account for the now firmly established rise in velocity dispersion
with age, which include spiral scattering, giant molecular cloud (GMC)
scattering, SF in turbulent gas, and the influence of tidal
interactions.  It is likely that all these mechanisms, and perhaps
others, contribute to the general increase of dispersion with age, but
he preferred the theory that it was largely driven by spiral
scattering, as was endorsed by \citet{Mack19}.  Note that resonant
scattering by spirals can drive up the in-plane components of random
motion, but has little effect on the vertical component; therefore a
population of GMCs is needed to redirect the in-plane peculiar
velocities into the vertical direction.

The ratio of vertical to radial dispersions $\ltsim 0.5$ ({\bf
  Figure~\ref{fig.AVRs}c}) among the low $[\alpha/{\rm Fe}]$ stars at,
and interior to, the Solar circle is surprisingly low, since cloud
scattering should quickly lead to a value of $\sim 0.6$, as predicted
by \citet{Ida93} for a flat rotation curve.  Their prediction, which
took proper account of distant scattering and is in agreement with
numerical results \citep{Vill85, ShId99, HaFl02, Sell08}, is that the
equilibrium axis ratio depends weakly on the local slope of the
rotation curve, with smaller values for declining, and larger for
rising, $V_c$ with radius.

Thus flatter values reported in {\bf Figure~\ref{fig.AVRs}c} could
perhaps be an indicator of a declining rotation curve, although
\citet{Ida93} expect a ratio as low as $\sim 0.4$ only when the
circular speed declines in a Keplerian fashion, which seems unlikely.
It is also possible that cloud scattering is inefficient such that the
expected equilibrium ratio has not been established, which would
require very few GMCs in the disk of the Milky Way.  However, the
ratios reported by \citet{Mack19} are lower than those found in
previous work \citep[\eg][]{Holm09} and, since their survey ranges
over greater distances from the Sun, could result from overestimates
of $\sigma_R$ if the second moment of the stellar velocity
distribution includes spatial variations of non-circular streaming
motions due to the bar and spirals.

The rounder shape of the ellipsoid at larger radii, indicated by the
red points in {\bf Figure~\ref{fig.AVRs}c}, is inconsistent with any
predicted slope of the rotation curve, and is likely due to satellite
heating, as also proposed by \citet{Mack19}.

Finally, the dispersions of the older, high $[\alpha/{\rm Fe}]$ stars
do not appear to change with age, which could merely reflect uncertain
age estimates.  Nevertheless, it is clear these stars constitute a
kinematically distinct population, and their random velocities are
thought to have been created by another mechanism, such as a minor
merger as the Milky Way accreted a satellite that thickened the then
disk, or stars forming in turbulent gas in the early stages of galaxy
assembly \citep[\eg][]{Genzel08}.

\subsection{Galaxy Formation}
\label{sec.galform}
Simulations of galaxy formation \citep[\eg][]{Vogels20} are proceeding
apace and the resulting galaxy models have a more authentic appearance
with almost every new paper, although quantitative differences from
the properties of real galaxies remain.  An attempt to survey this
progress would quickly become out-of-date, and would anyway be outside
the realm of this review.  Here we confine ourselves to a few remarks
related to spirals within the thin disks of the model galaxies that
constitute a challenge to the simulators.

A thin disk component in the model galaxies of a few years ago could
be recognized only among the very young star particles and gas.  A
clear example was given in Fig.~1 of \citet[][p3]{Bird13}, which
separated a simulated galaxy model at the present day into a number of
stellar ``age cohorts'' and presented face-on and edge-on projected
densities of each cohort.  Only the youngest cohort, those stars that
had formed within 0.5~Gyr of the moment of analysis, were in a thin
layer and manifested clear spiral patterns.  The surface density
profiles and thicknesses of each separate cohort were quantified in
Fig.~2 of their paper, and the radial and vertical velocity
dispersions in their Fig.~4.  The next youngest cohort having ages in
the range of 0.5~Gyr to 5.4~Gyr, had a much greater vertical thickness
and weak spiral patterns; no significant spirals could be discerned in
the still older and hotter cohorts.  The surface density of the
youngest cohort was just a few percent of the total projected density
and, consistent with swing-amplification theory (\S\ref{sec.swamp}),
supported multi-armed spirals in a low-mass cool disk.  The weaker
spirals in the second oldest, and more massive, cohort had lower
rotational symmetry, as theory would predict, but the greater velocity
dispersions and thickness inhibited strong patterns.  Thus the low
mass and limited age range of the thin disk in their model was at
variance with what we know of the thin disk in the Milky Way
\citep{BHGe16}, and nature of the spirals in their model bear little
resemblance to those in most galaxies (\S\ref{sec.obs}).  The fraction
of the disk mass in a thin component in the very recent FIRE-2
(Feedback in Realistic Environments) models \citep{YuBu21} reached
$\sim 50$\% in a few cases, but this is still not as large as it
should be.

The consequences of disks that are too hot and thick in the simulated
galaxies are clear. Spirals patterns tend to be unrealistically weak
and/or multi-armed, which has two consequences: (1) radial migration
is reduced, as reported by \citet{Bird13} and by \citet{Avil18}, for
the reasons given in \S\ref{sec.LMD}.  (2) Lindblad resonance
scattering plays a lesser role than it should in disk heating
(eq.~\ref{eq.Lchange}), although other mechanisms have clearly created
rather too much random motion.  Furthermore, the peculiar velocities
of the disk stars in real galaxies are re-oriented by scattering off
molecular clouds, whereas in galaxy formation simulations collisional
relaxation due to supermassive particles \citep[\eg][]{Sell14, Ludl21}
will have the same effect for the wrong reason!

\section{SUMMARY AND CONCLUSIONS}
The ubiquity of spiral patterns in the stellar disks of galaxies
requires them to result from self-excited instabilities within the
disk.  Other mechanisms, such as bars and tidal encounters, may well
drive spiral responses in specific cases, but we concluded in
\S\ref{sec.driven} that such external driving could not account for
all, or perhaps even most, spiral patterns.

The self-excitation mechanism in simulations of isolated, unbarred
disk galaxy models is now established to be a recurrent cycle of
groove modes (\S\ref{sec.best_theory}).  Individual modes, which have
constant pattern speed at all radii, grow and decay, with each having
significant amplitude for just a few turns at its corotation radius,
while new instabilities develop to maintain spiral activity.  However,
the superposition of several co-existing modes causes the spiral
appearance to change rapidly and the arms to appear to wind up over
time.  We argue that other theories have weaknesses
(\S\ref{sec.others}), and propose that a groove-mode cycle could be
responsible for spirals in real galaxies, as well as in simulations.
Observations (\S\ref{sec.obs}) suggest that swing amplification
(\S\ref{sec.responses}) plays a role in spiral formation, which is a
mechanism at the root of most theories, while evidence specific to a
recurrent cycle of groove modes is hard to obtain.  We have only hints
from the {\it Gaia} DR2 data that the distribution of disk stars in
the Milky Way (\S\ref{sec.tests}) manifests some of the features
expected from a groove-mode cycle.

The recurring patterns cause a secular increase in the random motions
of stars in the disk, reducing its responsiveness to subsequent
instabilities, and spiral activity in a purely stellar disk must fade
over time.  However, spiral activity can be maintained indefinitely if
the disk has even a modest fraction of gas, since gas clouds are able
to maintain a low velocity dispersion through dissipative collisions,
and form stars sharing similar kinematic properties, thereby
maintaining the responsiveness of the disk (\S\ref{sec.cooling}).

Spiral activity is a major driver of secular evolution in disk
galaxies.  It churns the disk stars, causing a radial diffusion that
flattens metallicity gradients (\S\ref{sec.rm}).  It also erases
density features in the disk (\S\ref{sec.flatten}), implying that the
smoothness of density profiles and of rotation curves need not be
properties that are required of galaxy formation.  The scattering of
stars at Lindblad resonances causes a secular rise in the in-plane
components of random motions, which can be scattered by GMCs into the
vertical direction.  These processes must contribute to the observed
increase in random motions of disk stars with age within the Milky Way
(\S\ref{sec.avr}) and in other galaxies.

Simulations of sub-maximum disks do not manifest spiral patterns that
are typical of most galaxies (\S\ref{sec.prefm}), and their multi-arm
features do not capture the full spiral-driven evolution of disk
galaxies (\S\ref{sec.LMD}).  These shortcomings are shared by galaxy
formation simulations that have not yet succeeded in creating thin
disks that are cool and massive enough to support realistic spiral
patterns (\S\ref{sec.galform}).

While we have reviewed the steady progress that has been made in the
development of our understanding of disk galaxy dynamics, we look
forward to more and better observational data to test these ideas (\S
\ref{sec:obssummary}).  Also, a number of outstanding theoretical
issues remain to be settled, which include the following:
\begin{enumerate}
\item Foremost is the problem of the stability of disks having gently
  rising rotation curves (\S\ref{sec.barstab}).  Despite years of
  effort, we have no satisfactory explanation for the absence of bars
  in such galaxies that often seem to manifest two-arm spiral
  patterns.
\item We have only a few specific models of spirals being tidally
  driven, and we need to know the mass range and orbit parameters of
  encounters with companion galaxies that can excite a spiral response
  without triggering a bar (\S\ref{sec.driven}).
\item We still lack compelling evidence that the recurrent cycle of
  groove modes, which has been identified as the mechanism for spiral
  generation in simulations (\S\ref{sec.best_theory}), also works in
  galaxies.  The later releases of {\it Gaia} data may yield stronger
  evidence in the Milky Way, but additional evidence from external
  galaxies would be highly desirable.
\item Transient spiral instabilities in heavy disks drive large-scale
  turbulence in the gas component in galaxies, which should strongly
  enhance magnetic field growth.  However, no direct tests of this
  prediction have yet been made (\S\ref{sec.magfld}).
\item Our discussion of the origin and effects of spiral patterns has
  concentrated on isolated galaxies, which present the most compelling
  need for a theoretical explanation.  We find on-going cosmological
  gas infall is needed to maintain self-excited spiral activity
  (\S\ref{sec.cooling}) and recognize that hierarchical clustering
  drives some spirals by tidal encounters (\S\ref{sec.driven}).  But
  the numerically challenging simulations of fully cosmological galaxy
  formation have not yet created cool stellar disks massive enough to
  support bi-symmetric spirals that are common in the nearby universe
  (\S\ref{sec.galform}).  Once that is achieved, we will be able to
  test whether the spiral dynamics discussed in this review applies in
  the full cosmological context.
\end{enumerate}

%Disclosure
\section*{DISCLOSURE STATEMENT}
The authors are not aware of any affiliations, memberships, funding,
or financial holdings that might be perceived as affecting the
objectivity of this review.

% Acknowledgements
\section*{ACKNOWLEDGMENTS}
We thank Scott Tremaine, John Kormendy, and the editor, Rob Kennicutt,
for many highly valuable comments on earlier drafts of this review.
Other comments from Ray Carlberg, Agris Kalnajs, and Igor Pasha were
also helpful.  JAS acknowledges the continuing hospitality of Steward
Observatory.

\def\etal{et al.}
\def\skip#1{ et al.}
\def\aap{{\it Astron. Astrophys.}}
\def\aaps{{\it Astron.\ \& Ap.\ Supp.}}
\def\acta{{\it Acta Astron.}}
\def\aj{{\it AJ}}
\def\an{{\it Astron. Nach.}}
\def\apj{{\it Ap.\ J.}}
\def\apjl{{\it Ap.\ J. Lett.}}
\def\apjs{{\it Ap.\ J. Suppl.}}
\def\aplc{{\it Ap.\ Lett.\ \& Comm.}}
\def\apss{{\it Ap.\ Sp.\ Sci.}}
\def\apssp{{\it Ap.\ Sp.\ Sci.\ Proc.}}
\def\araa{{\it Ann.\ Rev.\ Astron.\ Ap.}}
\def\arnps{{\it Annu.\ Rev.\ Nucl.\ Part.\ Sci.}}
\def\astronl{{\it Astron.\ Lett.}}
\def\astronr{{\it Astron.\ Rep.}}
\def\azh{{\it Ap.\ Zh.}}
\def\baas{{\it BAAS}}
\def\ban{{\it Bull.\ astr.\ Inst.\ Netherlands}}
\def\Cambridge{(Cambridge: Cambridge University Press)}
\def\cap{{\it Comments on Astrophysics}}
\def\fcp{{\it Fund.\ Cosmic Phys.}}
\def\gafd{{\it Geophys.\ Ap.\ Fluid Dyn.}}
\def\ica{{\it Icarus}}
\def\jcop{{\it J. Comp.\ Phys.}}
\def\jgr{{\it J. Geophys.\ Res.}}
\def\Kluwer{(Dordrecht: Kluwer)}
\def\lrr{{Liv.\ Rev.\ Rel.}}
\def\Messenger{{\it ESO Messenger}}
\def\mnras{{\it MNRAS}}
\def\nat{{\it Nature}}
\def\nast{{\it New Astron.}}
\def\obs{{\it Observatory}}
\def\pasa{{\it PAS Australia}}
\def\pasj{{\it PASJ}}
\def\pasp{{\it PASP}}
\def\pf{{\it Phys.\ Fluids}}
\def\PhD{{\it PhD thesis}}
\def\phya{{\it Physica\/} A}
\def\physcr{{\it Physica Scripta}}
\def\physl{{\it Phys.\ Lett.}}
\def\pnas{{\it Proc.\ Nat.\ Acad.\ Sci.\ (USA)}}
\def\prep{{\it Phys.\ Reports}}
\def\prev{{\it Phys.\ Rev.}}
\def\prd{{\it Phys.\ Rev.\ D}}
\def\prl{{\it Phys.\ Rev.\ Lett.}}
\def\pss{{\it Planet Sp.\ Sci.}}
\def\ptl{{\it Phil.\ Trans.\ R.\ Soc.\ London}}
\def\ptp{{\it Prog.\ Th.\ Phys.}}
\def\raa{{\it Res. Astr. \& Ap}}
\def\Reidel{(Dordrecht: Reidel)}
\def\rma{{\it Rev.\ Mod.\ Astr.}}
\def\rmp{{\it Rev.\ Mod.\ Phys.}}
\def\rnc{{\it Riv.\ Nuovo Cim.}}
\def\rpp{{\it Rep.\ Prog.\ Phys.}}
\def\sci{{\it Science}}
\def\sovast{{\it Sov.\ Astron.}}
\def\sovastl{{\it Soviet Ast.\ Lett.}}
\def\vistas{{\it Vistas Astron.}}

\end{document}